\documentclass[notitlepage,prd,showpacs,superscriptaddress,nofootinbib,floatfix,showkeys,10pt]{revtex4-1}
\usepackage{graphicx}
\usepackage{amsmath}
\usepackage{bm}
\usepackage{yhmath}
\usepackage{mathtools}
\usepackage{wasysym}
\usepackage{subfigure}
\usepackage{xcolor}
\definecolor{CherryRed}{rgb}{.65,0,.2}
\definecolor{RubyRed}{rgb}{.88,0.07,.3}
\definecolor{CralRed}{rgb}{1,0.25,.25}
\definecolor{CobaltBlue}{rgb}{0,0.28,.67}
\definecolor{RoyalBlue}{rgb}{0.25,0.41,.88}
\definecolor{EmeraldGreen}{rgb}{0.31,0.78,.47}
\usepackage{cases}
\usepackage[colorlinks=true, linkcolor=RubyRed,citecolor=CralRed,urlcolor=RubyRed,linkcolor=RubyRed]{hyperref}
\usepackage{subfigure}
\usepackage{times}
\usepackage{booktabs,bm}
\usepackage{slashed}
\usepackage{amsfonts,amssymb,stmaryrd,latexsym,amsmath}
\usepackage{textcomp}
\usepackage{multirow}
\usepackage{cancel}
\usepackage{array}
\usepackage{subcaption}
\usepackage{hyperref}
\usepackage{pgf}
\usepackage{orcidlink}
\usepackage{overpic}
\usepackage{graphicx}

\def\SU3{{\text{SU(3)}_{\rm F}}}

\def \pcs4338{{P_{\psi s}^\Lambda(4338)^0}}

\begin{document}
	
	\title{\textcolor{CobaltBlue}{XXXXXXXXX}}

	\date{\today}

	
\title{Phenomenology of Hypothetical Single-Top Hadronic States
}

\author{Z.~Rajabi Najjar$^{a}$\orcidlink{0009-0002-2690-334X}}
	\email{rajabinajar8361@ut.ac.ir }
\affiliation{Department of Physics, University of Tehran, North Karegar Avenue, Tehran 14395-547, Iran}
\author{M. Ahmadi\orcidlink{0009-0006-4046-2121}}
\email{masoumehahmadi@ut.ac.ir}
\affiliation{Department of Physics, University of Tehran, North Karegar Avenue, Tehran 14395-547, Iran}
\author{K. Azizi\orcidlink{0000-0003-3741-2167}}
\email{kazem.azizi@ut.ac.ir}
\thanks{Corresponding author}
\affiliation{Department of Physics, University of Tehran, North Karegar Avenue, Tehran
14395-547, Iran}
\affiliation{Department of Physics,  Faculty of Engineering and Natural Sciences,  Dogus University, Dudullu-\"{U}mraniye, 34775
Istanbul,  T\"{u}rkiye}

\date{\today}
	
\preprint{}
	
\begin{abstract}
We present a comprehensive theoretical study of the masses of possible baryonic and mesonic configurations containing a single top quark. Our analysis includes the baryons $\Lambda_t$, $\Xi_t$, $\Sigma_t$, $\Xi'_t$, $\Omega_t$, $\Omega_{tcc}$, and $\Omega_{tbb}$, together with the pseudoscalar and vector mesons $T_{t\bar n}^{\mathrm{Ps}}$, $T_{t\bar n}^{\mathrm{V}}$, $T_{t\bar s}^{\mathrm{Ps}}$, $T_{t\bar s}^{\mathrm{V}}$, $T_{t\bar c}^{\mathrm{Ps}}$, $T_{t\bar c}^{\mathrm{V}}$, $T_{t\bar b}^{\mathrm{Ps}}$, and $T_{t\bar b}^{\mathrm{V}}$. Motivated in part by recent experimental indications of a pseudoscalar enhancement near the $t\bar t$ threshold reported by the CMS and ATLAS collaborations, this study is carried out within the framework of two-point QCD sum rules to determine the corresponding ground-state masses by including perturbative contributions and nonperturbative condensates up to dimension eight. For several channels, including the $\Lambda_t$, $\Xi_t$, $\Sigma_t$, $T_{t\bar b}^{\mathrm{Ps}}$, and $T_{t\bar b}^{\mathrm{V}}$ states, the extracted central masses lie slightly below the corresponding sums of constituent quark masses, which may indicate nontrivial binding dynamics or near-threshold multiquark configurations within the uncertainties of the method. Moreover, when the full theoretical uncertainties are taken into account in a conservative manner, a larger subset of the investigated states exhibits a consistent tendency toward weak binding behavior, suggesting that the possibility of loosely bound configurations cannot be excluded for most of the considered baryonic and mesonic channels. These results provide useful first-principles theoretical benchmarks for possible top-containing hadronic systems, which may support future searches at the LHC, along with sensitivity analyses for next-generation facilities such as the FCC.
		
\end{abstract}
	
\keywords{}
	
	
\maketitle
	
\renewcommand{\thefootnote}{\#\arabic{footnote}}
\setcounter{footnote}{0}

	\maketitle
	\renewcommand{\thefootnote}{\#\arabic{footnote}}
	\setcounter{footnote}{0}


\section {Introduction}\label{sec:one}
As the heaviest known elementary particle, the top quark, with a mass of approximately $173~\text{GeV}$, occupies a distinctive role within the Standard Model. In contrast to lighter quarks, its extremely short lifetime, of the order of $5 \times 10^{-25}\,\text{s}$, prevents it from undergoing hadronization prior to decay. This unique feature poses considerable theoretical and experimental challenges, while at the same time providing a valuable framework for precision studies in hadron physics  \cite{CDF:1995wbb,D0:1995jca}. Among all quark flavors, the top quark is distinguished by its exceptionally short lifetime \cite{Bigi:1986jk,Shibata:2008sy}.
Beyond serving as a powerful test of the Standard Model, the top quark is especially significant because its fundamental properties offer a direct window into potential new physics beyond the Standard Model \cite{Schrempp:1996fb,Atwood:2000tu,Kehoe:2007px,Plehn:2011tg,Wagner:2005jh,Quadt:2006dqn,Galtieri:2011yd,Aguilar-Saavedra:2014kpa,Hagiwara:2016rdv,Chen:2016zbz,Agashe:2016bok}. Despite the extremely short lifetime of the top quark, which prevents the formation of a stable bound state with its antiparticle, $t\bar{t}$ production near threshold is nevertheless described by nonrelativistic QCD dynamics. Due to the extremely short lifetime of the top quark, it has long been generally understood that there is insufficient time for the formation of a true quarkonium-like bound state, a view consistently supported in standard particle physics literature \cite{Kuhn:1987ty,Barger:1987xg,Fadin:1990wx,Sumino:1997ve,Penin:2005eu,Hagiwara:2008df,Kiyo:2008bv,Beneke:2015kwa,Fuks:2021xje,Akbar:2024brg,Wang:2024hzd,Jiang:2024fyw,Fuks:2024yjj,Fadin:1987wz,Strassler:1990nw,Fadin:1994pj,Hoang:2000yr,Sumino:2010bv,Kawabata:2016aya,Reuter:2018rbq,Aguilar-Saavedra:2024mnm,Garzelli:2024uhe,Jafari:2025rmm,Francener:2025tor,dEnterria:2025ecx}. However, this traditional understanding has recently been strongly questioned by measurements reported by the CMS and ATLAS Collaborations, which observe an enhancement near the $t\bar t$ mass threshold. This feature has been interpreted as a possible indication of the formation of a pseudoscalar toponium state \cite{CMS:2025kzt,ATLAS:2025mvr}. A notable quantum feature of \(t\bar{t}\) production near threshold is the strong spin correlation between the quarks. This behavior may originate from near-threshold dynamics, which can affect measurable observables despite the extremely short lifetimes of the top and antitop quarks. While the existence of a toponium state and its properties had been considered in earlier work, decay-channel measurements with a significance above \(5\sigma\) \cite{CMS:2025kzt} have recently stimulated considerable theoretical activity \cite{Fuks:2025toq,Biello:2025idh,Matsuoka:2025jgm,Sjostrand:2025qez,Fuks:2025wtq,Goncalves:2025hyx,Zhu:2025ezg,Zhang:2025fdp,Lopez:2025kog,Thompson:2025cgp,Zhang:2025xxd,Shao:2025dzw,Bai:2025buy,Ellis:2025nkm,LeYaouanc:2025mpk,Xiong:2025iwg,Fu:2025yft,Fu:2025zxb,Fuks:2025sxu,Afik:2025ejh,Luo:2025psq,Jolly:2026gpe,Garzelli:2026jxh,CMS:2026lsc,Najjar:2025bby}. 

 The investigation of the internal structure and the prediction of the mass spectra of baryons and mesons containing one, two, or three heavy quarks constitute an important area of research in hadronic physics. In particular, hadrons with a single heavy quark have been extensively studied and are now well established both theoretically and experimentally. On the theoretical side, relativistic quark models, Heavy Quark Effective Theory (HQET), and lattice QCD have been successfully employed to describe the mass spectra, decay properties, and internal structure of these states~\cite{Patel:2025gbw,Li:2023gbo,Li:2025frt,Gayer:2024akw,Kaneko:2023kxx,Luo:2025sns,Ni:2023lvx,Fabiano:1997xh}. On the experimental side, significant progress has been achieved in the observation and precision spectroscopy of hadrons containing a single heavy quark, particularly in the charm and bottom sectors. A large number of singly heavy mesons and baryons, such as $\Lambda_c$, $\Xi_c$, $\Omega_c$, $\Lambda_b$, and $\Xi_b$, have been observed, and their properties have been measured with high precision in high-energy experiments. In particular, the LHCb Collaboration \cite{LHCb:2017uwr,LHCb:2020iby} has reported numerous new excited states and provided detailed measurements of their masses and decay widths, and in several cases also their quantum numbers. Complementary measurements from the Belle \cite{Belle:2013jfq} and BaBar \cite{BaBar:2010zpy} experiments have further confirmed and refined the spectroscopy of charm hadrons, providing crucial cross-checks of the LHCb results. These successes make singly heavy hadrons a crucial testing ground for nonperturbative QCD and provide a solid foundation for extending theoretical models to systems with multiple heavy quarks or even hypothetical top-containing hadrons.

 Recent renewed interest in the possible formation of toponium near the top--antitop threshold has stimulated further investigations into whether the Standard Model may also accommodate hadronic systems containing top quarks. In particular, the study of mesons and baryons containing a single top quark is especially appealing, as such systems could provide a unique probe of QCD dynamics at energy scales that have not been explored before. Although the very short lifetime of the top quark has traditionally been considered a major obstacle to hadron formation, recent theoretical and phenomenological studies suggest that bound-state effects and formation times may be compatible with the top-quark lifetime in certain kinematic regimes. Owing to the exceptionally short lifetime of the top quark, the formation of top-containing hadrons is expected to be strongly suppressed, rendering their experimental observation highly challenging. Nevertheless, if such states can form under suitable dynamical conditions, the mass predictions presented in this work may provide useful guidance for future theoretical investigations, lattice QCD studies, and experimental searches near the top--antitop threshold \cite{Garzelli:2024uhe,Fuks:2024yjj,Fuks:2025toq,Najjar:2025bby}. 

 In this work, we investigate the masses of baryons and mesons that contain a single-top quark using the framework of QCD sum rules. This method, grounded in the QCD Lagrangian, provides a robust theoretical tool for predicting hadronic properties, particularly in systems involving heavy quarks. Its reliability has been well established through successful applications to heavy hadrons, yielding results in good agreement with experimental data \cite{Aliev:2009jt,Aliev:2010uy,Aliev:2012ru,Agaev:2016mjb,Azizi:2016dhy}. Within this framework, hadron masses are determined from two-point correlation functions constructed with appropriately chosen interpolating currents. The calculation relies on the operator product expansion (OPE), supplemented by the Borel transformation to enhance ground-state contributions, and employs continuum subtraction based on the quark--hadron duality assumption.

The paper is organized as follows. Section \ref{sec:two} presents the QCD sum rules formalism and describes the computational framework in detail. In Section \ref{sec:three}, we report the mass predictions for baryons and mesons containing a single top quark, along with a comparative analysis of the results. Finally, Section \ref{sec:four} provides a summary of the main findings and discusses the implications of the present work.

\section {Mass evaluation within the framework of QCD sum rules}\label{sec:two}

In this work, we employ the QCD sum rules framework to investigate the masses of baryons and mesons in both pseudoscalar and vector states containing a single-top quark. This approach offers a powerful and systematic way to connect hadronic properties with the underlying quark and gluon degrees of freedom in QCD.
The present analysis is carried out in two main steps. In the first, the correlation function is expressed on the phenomenological side in terms of hadronic parameters, including the mass and decay constant of the state under consideration. In the second step, the same correlation function is evaluated within QCD using the OPE, where nonperturbative effects are incorporated through quark, gluon and mixed condensates contributions. By constructing two representations of the same correlation function, namely the hadronic representation and the QCD representation, the two sides are related through a dispersion relation under the assumption of quark--hadron duality. This procedure leads to the formulation of QCD sum rules for the physical observables under consideration. The relevant parameters are then extracted by matching, in a consistent manner, the coefficients of the corresponding Lorentz structures obtained from both representations.

To derive the QCD sum rules for single-top baryons and mesons, one first introduces an appropriate  correlation function consistent with the quantum numbers of the state under investigation for each channel. The analysis starts with the construction of correlators formed from the time‑ordered products of the relevant interpolating currents. These correlators form the basis of the QCD sum rule formalism, as they establish a connection between hadronic observables and the fundamental quark–gluon dynamics of QCD. In general, the correlation function takes the form:

\begin{eqnarray}
\Pi_{\overline{\mathbf{3}}}(q)
&=& i \int d^{4}x \, e^{iq \cdot x}
\langle 0 | \mathcal{T}
\left\{
J_{\overline{\mathbf{3}}}(x)\,
\bar{J}_{\overline{\mathbf{3}}}(0)
\right\}
| 0 \rangle,\nonumber\\
\Pi_ { 6}(q)&=& i \int d^{4}x \, e^{iq \cdot x} \, \langle 0 | \mathcal{T} \{ J_{6}(x)\,\bar{J}_6(0) \} | 0 \rangle,\nonumber\\
\Pi_ { B_Q}(q)&= &i \int d^{4}x \, e^{iq \cdot x} \, \langle 0 | \mathcal{T} \{ J_{ B_Q}(x)\,\bar{J}_{ B_Q}(0) \} | 0 \rangle.
\label{QCDD1}
\end{eqnarray}
In this expression, \( J_{\overline{\mathbf{3}}} \), \( J_{\mathbf{6}} \), and \( J_{\mathbf{B_Q}} \) denote the interpolating currents associated with single-heavy baryon states belonging to the flavor anti-triplet (\(\overline{\mathbf{3}}\)) and sextet (\(\mathbf{6}\)) representations, as well as triply heavy baryon states, respectively. Here, \( \mathcal{T} \) denotes the time-ordering operator, and \( q \) is the four-momentum carried by the hadronic state. To obtain the QCD sum rules for the systems under consideration, it is necessary to introduce appropriate interpolating currents for each relevant channel. The corresponding expressions are given below \cite{Bagan:1991sc}:

\begin{align}\label{QCDD2}
J_{\overline{\mathbf{3}}}&= \frac{1}{\sqrt{6}} \epsilon_{abc} \sum_{l=1}^2
\Bigg\{
2\Big( q_1^{T,a}(x) C \Gamma_{1}^{l} q_2^{b}(x)\Big)\Gamma_{2}^{l} t^c(x) + \Big( q_1^{T,a}(x) C \Gamma_{1}^{l} t^b(x)\Big)\Gamma_{2}^{l} q_2^{c}(x)+ \Big( t^{T,a}(x) C \Gamma_{1}^{l} q_2^b(x)\Big)\Gamma_{2}^{l} q_1^{c}(x)
\Bigg\},\nonumber\\
J_{\mathbf{6}} &= - \frac{1}{\sqrt{2}} \epsilon_{abc} \sum_{l=1}^2
\Bigg\{
\Big( q_1^{T,a}(x) C \Gamma_{1}^{l} t^{b}(x)\Big)\Gamma_{2}^{l} q_2^c(x)
- \Big( t^{T,a}(x) C \Gamma_{1}^{l} q_2^b(x)\Big)\Gamma_{2}^{l} q_1^{c}(x)
\Bigg\},\nonumber\\
		J_{\mathbf{B_Q}}&=2\epsilon_{abc}\Big\{\Big(Q^{T,a}(x)Ct^{b}(x)\Big)\gamma_{5}Q^c(x)+
		\beta\Big(Q^{T,a}(x)C\gamma_{5}t^{b}(x)\Big)Q^c (x)\Big\}.
	\end{align}

Here, $\epsilon_{abc}$ denotes the Levi-Civita tensor in color space with indices $a$, $b$, and $c$, and $C$ represents the charge conjugation operator. The Dirac structures are defined as $\Gamma_1^1 = 1$, $\Gamma_1^2 = \Gamma_2^1 = \gamma^5$, and $\Gamma_2^2 = \beta$, where $\beta$ is an arbitrary mixing parameter. In particular, the choice $ \beta=-1$ corresponds to the well-known Ioffe currents. The fields $q_1$ and $q_2$ denote light-quark fields, while $t$ represents the top-quark field. The field $Q$ corresponds to the charm- or bottom-quark field.
To construct the QCD sum rules for the mesonic states, namely the pseudoscalar and vector channels, it is necessary to introduce an appropriate correlation function to analyze separately on the physical and the QCD sides:
\begin{align}
\Pi^{Ps}(q) &= i \int d^{4}x\, e^{iqx}\, \langle 0|\mathcal{T}\{J^{Ps}(x)\, J^{\dagger Ps}(0)\}|0\rangle, \notag \\
\Pi_{\mu\nu}^{V}(q) &= i \int d^{4}x\, e^{iqx}\, \langle 0|\mathcal{T}\{J_\mu^{V}(x)\, J_\nu^{\dagger V}(0)\}|0\rangle.
\label{QCDD3}
\end{align} 
 Here, $J^{Ps}$, $J^{V}$ represent the interpolating currents for the pseudoscalar and vector systems respectively as follows:

\begin{eqnarray}
J^{Ps} (x)&=&\overline{q}^{ a}(\overline{Q}^{ a})(x)\gamma_5t^{ b}(x),\nonumber\\	
J^{V}_\mu (x)&=&\overline{q}^{ a}(\overline{Q}^{ a})(x)\gamma_\mu t^{ b}(x).
\label{QCDD4}
\end{eqnarray}
The fields $q$ represent light quarks.
To proceed, the two-point correlation function constructed from the corresponding interpolating currents is analyzed in two complementary representations for each case: the phenomenological (hadronic) representation, which encodes the contributions of physical intermediate states, and the QCD representation, which is evaluated within the framework of OPE.

\subsection {The hadronic (phenomenological) representation}

In the QCD sum rule approach, the phenomenological representation of the correlation function is constructed by saturating it with a complete set of hadronic states carrying the same quantum numbers as the interpolating current. After carrying out the spacetime integration, the ground-state contribution is isolated from the effects of excited resonances and the continuum spectrum. Consequently, the correlation function for the baryonic system under consideration can be expressed in its hadronic form as follows:

\begin{equation}
\Pi^{\text{Phys}}(q) =
\frac{\langle 0 | J | B(q,s) \rangle
\langle B(q,s) | \bar{J} | 0 \rangle}
{m_B^{2} - q^{2}}
+ \cdots.
\end{equation}

The state \( |B(q,s)\rangle \) is identified with the single-particle ground-state baryon, where \(s\) denotes its spin projection. Contributions from higher excited resonances and the continuum spectrum are represented by the ellipsis, while the baryon mass is denoted by \( m_B \).

Similarly, the correlation function for the pseudoscalar meson states, \(T^{Ps}\), can be expressed in its hadronic form as follows:

\begin{eqnarray}
\Pi^{\mathrm{Phys}}
&=&
\frac{{\langle}0| J^{Ps} | T^{Ps}(q,s)\rangle
\langle T^{Ps}(q,s)| J^{\dagger Ps}|0\rangle}
{m_{Ps}^2-q^2}
+\cdots.
\label{Eq:cor:Phys1}
\end{eqnarray}

Also, for the vector meson states, \(T^{V}\):

\begin{eqnarray}
\Pi^{\mathrm{Phys}}_{\mu\nu}
&=&
\frac{{\langle}0| J^{V}_\mu | T^{V}(q,r)\rangle
\langle T^{V}(q,r)| J^{\dagger V}_\nu|0\rangle}
{m_{V}^2-q^2}
+\cdots.
\label{Eq:cor:Phys2}
\end{eqnarray}

The quantities \( m_{Ps} \) and \( m_V \) correspond to the masses of the pseudoscalar and vector mesons, respectively, while \( |T^{Ps}(q,s)\rangle \) and \( |T^{V}(q,r)\rangle \) denote the corresponding states. Here, \(r\) labels the polarization state of the vector meson. For a complete analysis, the matrix elements of these currents between the vacuum and the relevant hadronic states must be determined, covering both mesonic and baryonic channels. These matrix elements are parameterized in terms of the decay constants \( f_{Ps} \), \( f_V \), the baryonic residue \( \lambda \), and the corresponding masses as

\begin{equation}
\begin{aligned}
\langle 0 | J | B(q,s) \rangle
&= \lambda\, u(q,s), \\
\langle 0 | J^{Ps} | T^{Ps}(q,s)\rangle
&= f_{Ps}\,\frac{m_{Ps}^2}{m_t+m_{q/Q}}, \\
\langle 0 | J^V_\mu | T^V(q,r)\rangle
&= f_V\, m_V\, \varepsilon_\mu^{(r)} .
\end{aligned}
\end{equation}

Here, \(u(q,s)\) is the Dirac spinor of the baryon with spin projection \(s\), and \(\varepsilon_\mu^{(r)}\) is the polarization four-vector corresponding to the polarization state \(r\) of the vector meson. The polarization sum is given by

\begin{equation}
\sum_{r}
\varepsilon_{\mu}^{(r)*}(q)\,
\varepsilon_{\nu}^{(r)}(q)
=
-\left(
g_{\mu\nu}
-\frac{q_{\mu} q_{\nu}}{m_V^2}
\right),
\label{sum}
\end{equation}

while the spin sum reads

\begin{equation}
\sum_s u(q,s)\bar u(q,s)
=
(\slashed q + m_B).
\end{equation}

The phenomenological (physical) representations of the correlation functions corresponding to the baryonic states and the mesonic states \(T^{Ps}\) and \(T^{V}\) are obtained as follows:

\begin{align}\label{phen2}
\Pi^{\mathrm{Phys(B)}}(q) &=
\frac{\lambda^{2}}{m_{B}^{2}-q^{2}}
(\!\not\!{q} + m_{B}), \\
\Pi^{\mathrm{Phys(Ps)}} &=
\frac{f_{Ps}^2
\left(\frac{m_{Ps}^2}{m_t+m_{q/Q}}\right)^2}
{m_{Ps}^2-q^2}
+\cdots,
\nonumber\\[4pt]
\Pi_{\mu\nu}^{\mathrm{Phys(V)}} &=
\frac{f_{V}^2 m_{V}^2}
{m_{V}^2-q^2}
\left[
-g_{\mu\nu}
+\frac{q_{\mu}q_{\nu}}{m_{V}^2}
\right]
+\cdots.
\nonumber
\end{align}

Finally, the application of the Borel transformation with respect to \( q^2 \) efficiently suppresses the contributions of higher resonances and the continuum, thereby enhancing the dominance of the ground-state contribution. Consequently, the hadronic representation of the correlation functions for the particles under consideration is obtained as follows:

\begin{eqnarray}
\tilde{\Pi}^{\mathrm{Phys(Ps)}}(q)
&=&
f_{Ps}^2
\left(\frac{m_{Ps}^2}{m_t+m_{q/Q}}\right)^2
e^{-\frac{m^{2}_{Ps}}{M^{2}}}
+\cdots,
\nonumber\\
\tilde{\Pi}_{\mu \nu}^{\mathrm{Phys(V)}}(q)
&=&
f_{V}^2 m_{V}^2
e^{-\frac{m^{2}_V}{M^{2}}}
\left[
-g_{\mu\nu}
+\frac{q_{\mu}q_{\nu}}{m_{V}^2}
\right]
+\cdots,
\nonumber\\
\tilde{\Pi}^{\mathrm{Phys(B)}}(q)
&=&
\lambda^2
e^{-\frac{m^{2}_B}{M^{2}}}
(\!\not\!{q}+ m_B)
+\cdots.
\label{eq:CorFunBorel}
\end{eqnarray}

Here, \( \widetilde{\Pi}^{\mathrm{Phys}}(q) \) represents the Borel-transformed correlation function, while \( M^{2} \) denotes the corresponding Borel parameter.

\subsection {QCD representation}
To compute the QCD side, we start from the correlation functions given in Eq.~(\ref{QCDD1}) by inserting the interpolating currents of the corresponding baryons defined in Eq.~(\ref{QCDD2}). In a similar manner, the correlation function in Eq.~(\ref{QCDD3}) is obtained by employing the interpolating currents of the relevant mesons introduced in Eq.~(\ref{QCDD4}). The evaluation proceeds by performing all possible contractions of the quark fields via Wick’s theorem. The resulting quark operators can then be expressed in terms of the corresponding light- and heavy-quark propagators in coordinate space, whose explicit forms are given below \cite{Sundu:2018nxt}

\begin{eqnarray}
	   	&&S_{q}^{ab}(x)=i\delta _{ab}\frac{\slashed x}{2\pi ^{2}x^{4}}-\delta _{ab}%
	   	\frac{m_{q}}{4\pi ^{2}x^{2}}-\delta _{ab}\frac{\langle \overline{q}q\rangle
	   	}{12}+i\delta _{ab}\frac{\slashed xm_{q}\langle \overline{q}q\rangle }{48}%
	   	-\delta _{ab}\frac{x^{2}}{192}\langle \overline{q}g_{s}\sigma Gq\rangle
	   	\notag \\
	   	&&+i\delta _{ab}\frac{x^{2}\slashed xm_{q}}{1152}\langle \overline{q}%
	   	g_{s}\sigma Gq\rangle -i\frac{g_{s}G_{ab}^{\alpha \beta }}{32\pi ^{2}x^{2}}%
	   	\left[ \slashed x{\sigma _{\alpha \beta }+\sigma _{\alpha \beta }}\slashed x%
	   	\right] -i\delta _{ab}\frac{x^{2}\slashed xg_{s}^{2}\langle \overline{q}%
	   		q\rangle ^{2}}{7776}  \notag \\
	   	&&-\delta _{ab}\frac{x^{4}\langle \overline{q}q\rangle \langle
	   		g^2_{s}G^{2}\rangle }{27648}+\cdots ,  \label{eq:A1}
	   \end{eqnarray}%
	   and
	   \begin{eqnarray}
	   	&&S_{Q}^{ab}(x)=i\int \frac{d^{4}k}{(2\pi )^{4}}e^{-ikx}\Bigg \{\frac{\delta
	   		_{ab}\left( {\slashed k}+m_{Q}\right) }{k^{2}-m_{Q}^{2}}-\frac{%
	   		g_{s}G_{ab}^{\alpha \beta }}{4}\frac{\sigma _{\alpha \beta }\left( {\slashed %
	   			k}+m_{Q}\right) +\left( {\slashed k}+m_{Q}\right) \sigma _{\alpha \beta }}{%
	   		(k^{2}-m_{Q}^{2})^{2}}  \notag  \label{eq:A2} \\
	   	&&+\frac{g_{s}^{2}G^{2}}{12}\delta _{ab}m_{Q}\frac{k^{2}+m_{Q}{\slashed k}}{%
	   		(k^{2}-m_{Q}^{2})^{4}}+\frac{g_{s}^{3}G^{3}}{48}\delta _{ab}\frac{\left( {%
	   			\slashed k}+m_{Q}\right) }{(k^{2}-m_{Q}^{2})^{6}}\left[ {\slashed k}\left(
	   	k^{2}-3m_{Q}^{2}\right) +2m_{Q}\left( 2k^{2}-m_{Q}^{2}\right) \right] \left(
	   	{\slashed k}+m_{Q}\right) +\cdots \Bigg \}. \notag \\
	   	&&
	   \end{eqnarray}
Here, \( G_{\mu\nu} \) denotes the gluon field-strength tensor, whose color structure is decomposed as \( G^{\alpha\beta}_{ab} = G^{\alpha\beta}_{A} t^{A}_{ab} \), with \( t^{A} = \lambda^{A}/2 \). The gluonic operators are defined by \( G^{2} = G_{\alpha\beta}^{A} G_{A}^{\alpha\beta} \) and \( G^{3} = f^{ABC} G_{\alpha\beta}^{A} G^{B\,\beta\delta} G_{\delta}^{C\,\alpha} \). In this notation, \( \lambda^{A} \) are the Gell–Mann matrices generating the \( SU_{c}(3) \) color group, and \( f^{ABC} \) denote its totally antisymmetric structure constants, with \( A,B,C = 1,\ldots,8 \). Consequently, the correlation functions for baryonic and mesonic channels can be systematically expressed in terms of the corresponding heavy- and light-quark propagators as follows:

\begin{eqnarray}\label{antisymmetricq1difq2}
\Pi_{\overline{\textbf{3}}}^{\mathrm{QCD}}(q) &=& \frac{i}{6}\epsilon_{abc}\epsilon_{a'b'c'} \int d^4 x \, e^{iq\cdot x}
\sum_{l=1}^2 \sum_{k=1}^2 \Big\{ \Gamma^{l}_{2} \Big[ 2 S^{ca'}_{t}(x) \Gamma^{k}_{1}\widetilde{S}^{ab'}_{q_{1}}(x) \Gamma^{l}_{1} S^{bc'}_{q_{2}}(x) \Gamma^{k}_{2} \nonumber\\
&& + 2 S^{cb'}_{t}(x) \Gamma^{k}_{1}\widetilde{S}^{ba'}_{q_{2}}(x) \Gamma^{l}_{1} S^{ac'}_{q_{1}}(x) \Gamma^{k}_{2}
- S^{ca'}_{q_{2}}(x) \Gamma^{k}_{2}\widetilde{S}^{bb'}_{t}(x) \Gamma^{l}_{1} S^{ac'}_{q_{1}}(x) \Gamma^{k}_{1} \nonumber\\
&& - 2 S^{ca'}_{q_{2}}(x) \Gamma^{k}_{2}\widetilde{S}^{ab'}_{q_{1}}(x) \Gamma^{l}_{1} S^{bc'}_{t}(x) \Gamma^{k}_{1}
- S^{cb'}_{q_{1}}(x) \Gamma^{k}_{2}\widetilde{S}^{aa'}_{t}(x) \Gamma^{l}_{1} S^{bc'}_{q_{2}}(x) \Gamma^{k}_{1} \nonumber\\
&& - 2 S^{cb'}_{q_{1}}(x) \Gamma^{k}_{2}\widetilde{S}^{ba'}_{q_{2}}(x) \Gamma^{l}_{1} S^{ac'}_{t}(x) \Gamma^{k}_{1} \Big] \nonumber\\
&& - \Gamma^{k}_{2} \Big[ S^{cc'}_{q_{1}}(x) \Gamma^{l}_{2} \mathrm{Tr}(\Gamma^{k}_{1} S^{ab'}_{t}(x) \Gamma^{l}_{1}\widetilde{S}^{ba'}_{q_{2}}(x)) \nonumber\\
&& + S^{cc'}_{q_{2}}(x) \Gamma^{l}_{2} \mathrm{Tr}(\Gamma^{k}_{1} S^{ab'}_{q_{1}}(x)\Gamma^{l}_{1} \widetilde{S}^{ba'}_{t}(x))
+ 4 S^{cc'}_{t}(x) \Gamma^{l}_{2} \mathrm{Tr}(\Gamma^{k}_{1} S^{ab'}_{q_{1}}(x)\Gamma^{l}_{1} \widetilde{S}^{ba'}_{q_{2}}(x)) \Big] \Big\},
\end{eqnarray}

\begin{eqnarray}\label{symmetricq1difq2}
\Pi_{\textbf{6}}^{\mathrm{QCD}}(q) &=& -\frac{i}{2}\epsilon_{abc}\epsilon_{a'b'c'} \int d^4 x \, e^{iq\cdot x}
\sum_{l=1}^2 \sum_{k=1}^2 \Big\{ \Gamma^{l}_{2} \Big[ S^{ca'}_{q_{1}}(x) \Gamma^{k}_{1}\widetilde{S}^{ab'}_{t}(x) \Gamma^{l}_{1} S^{bc'}_{q_{2}}(x) \nonumber\\
&& + S^{cb'}_{q_{2}}(x) \Gamma^{k}_{1}\widetilde{S}^{ba'}_{t}(x) \Gamma^{l}_{1} S^{ac'}_{q_{1}}(x) \Big]\Gamma^{k}_{2}
+ \Gamma^{k}_{2} \Big[ S^{cc'}_{q_{1}}(x) \Gamma^{l}_{2} \mathrm{Tr}(\Gamma^{l}_{1} \widetilde{S}^{aa'}_{t}(x)\Gamma^{k}_{1}S^{bb'}_{q_{2}}(x)) \nonumber\\
&& + S^{cc'}_{q_{2}}(x) \Gamma^{l}_{2} \mathrm{Tr}(\Gamma^{l}_{1} \widetilde{S}^{aa'}_{q_{1}}(x)\Gamma^{k}_{1}S^{bb'}_{t}(x)) \Big] \Big\},
\end{eqnarray}
\begin{eqnarray}\label{eqQCD1}
\Pi^{QCD}_{B_Q}(q)&=&4i\epsilon_{abc}\epsilon_{a'b'c'}\int d^4x e^{i q x} \Big\{-\gamma_{5}
S^{c'b}_{Q}(x)\widetilde{S}^{b'a}_{t}(x)S^{a'c}_{Q}(x)\gamma_{5}+
\gamma_{5}S^{c'c}_{Q}(x)\gamma_{5}Tr\Big[S^{a'b}_{Q}(x)\widetilde{S}^{b'a}_{t}(x)\Big]  \notag \\
&+& \beta\Big( -\gamma_{5}S^{c'b}_{Q}(x)\gamma_{5}\widetilde{S}^{b'a}_{t}(x)S^{a'c}_{Q}(x)
-S^{c'b}_{Q}(x)\widetilde{S}^{b'a}_{t}(x)\gamma_{5}S^{a'c}_{Q}(x)\gamma_{5}
+\gamma_{5}S^{c'c}_{Q}(x)Tr\Big[S^{a'b}_{Q}(x)\gamma_{5}\widetilde{S}^{b'a}_{t}(x)\Big]  \notag \\
&+& S^{c'c}_{Q}(x)
\gamma_{5}Tr\Big[S^{a'b}_{Q}(x)\widetilde{S}^{b'a}_{t}(x)\gamma_{5}\Big]\Big) \notag \\
&+&
\beta^2\Big( -S^{c'b}_{Q}(x)\gamma_{5}\widetilde{S}^{b'a}_{t}(x)\gamma_{5}S^{a'c}_{Q}(x)+S^{c'c}_{Q}(x)
Tr\Big[S^{b'a}_{t}(x)\gamma_{5}\widetilde{S}^{a'b}_{Q}(x)\gamma_{5}\Big]
\Big)
\Big\},
\end{eqnarray}

\begin{eqnarray}
\label{QCDSide1}
\Pi^{\mathrm{Ps}} (q) &=& i \int d^4x
e^{iqx}    Tr \Big[\gamma_5
S_{q(Q)}^{ a a'} \gamma_5 S_t^{bb'}\Big], \nonumber \\
\Pi_{\mu\nu}^{\mathrm{V}} (q) &=& i  \int d^4x
e^{iqx}   Tr \Big[\gamma_\mu
S_{q(Q)}^{ a a'} \gamma_\nu S_t^{bb'}\Big].
\end{eqnarray}
The QCD representations for the baryonic channels are given in Eqs.~(\ref{antisymmetricq1difq2}), (\ref{symmetricq1difq2}), and (\ref{eqQCD1}) for the antisymmetric $\overline{\textbf{3}}$, symmetric $\textbf{6}$, and triply heavy configurations, respectively. For the mesonic channels, the corresponding correlation functions are given in Eq. (\ref{QCDSide1}). Here, $\widetilde{S}^{ij} = C S^{ij,T} C$ denotes the charge-conjugated quark propagator. 

 Following the substitution of the quark propagators into the correlation
functions, the Wick contractions of the quark fields are carried out
according to Wick's theorem. The resulting expressions are then simplified
by evaluating the Dirac and color traces, after which the relevant
Lorentz structures are isolated. Applying the Fourier transformation,
followed by the Borel transformation, the invariant amplitudes are cast
into a dispersion representation in terms of the corresponding spectral
densities $\rho_i(s)$ and subtraction terms $\Gamma_i(M^2)$. Finally,
the contributions from higher resonances and the continuum are suppressed
using the quark--hadron duality ansatz. Consequently, the QCD side can be
written for the baryons and mesons as

\begin{equation}\label{eqQCD4}
\begin{aligned}
\tilde\Pi^{\mathrm{QCD}}_{\!\not\!q}(s_0,M^2)
&=
\int_{s'}^{s_0} ds\, e^{-s/M^2}\rho_{\!\not\!q}(s)
+\Gamma_{\!\not\!q}(M^2), \\
\tilde\Pi^{\mathrm{QCD(Ps)}}(s_0,M^2)
&=
\int_{s'}^{s_0} ds\, e^{-s/M^2}\rho_{Ps}(s)
+\Gamma_{Ps}(M^2), \\
\tilde\Pi^{\mathrm{QCD(V)}}_{g_{\mu\nu}}(s_0,M^2)
&=
\int_{s'}^{s_0} ds\, e^{-s/M^2}
\rho_{g_{\mu\nu}}(s)
+\Gamma_{g_{\mu\nu}}(M^2).
\end{aligned}
\end{equation}
In this formulation, the parameter $s_0$ is introduced to characterize the threshold that separates the ground-state contribution from that of the continuum, while $s'$ represents the square of the sum of the quark masses of the hadron under study. On both the QCD and physical sides, the baryon and vector-meson states under consideration admit two independent Lorentz structures. These can be identified as $(\!\not\!q,I)$ and $(g_{\mu\nu},\, q_{\mu}q_{\nu})$. In the present analysis, the sum rules are extracted from the Lorentz structures $g_{\mu\nu}$ for the vector mesons and $\!\not\!q$ for the baryons, as these structures are found to provide good convergence of the operator product expansion (OPE) and stable numerical predictions.
The spectral distributions $\rho_i(s)$, with $ i= \!\not\!q, Ps, g_{\mu\nu}$,  are defined as:
		   	  \begin{equation}
\rho_i(s)=\frac{1}{\pi}\operatorname{Im}\tilde{\Pi}_i(q^2).
\end{equation}
 Furthermore, the functions $\Gamma_i(M^2)$ are real-valued  functions, directly calculated from the correlation functions.  The $\rho_i(s)$ and $\Gamma_i(M^2)$ functions together contain all perturbative and nonperturbative contributions arising from quark, gluon, and mixed condensates of mass dimension up to eight. Their expressions are lengthy, hence we do not present their explicit expressions here.
  
 The results obtained from the phenomenological and QCD representations are matched through dispersion relations by equating the coefficients of the same Lorentz structures. This matching procedure yields the QCD sum rule expressions for the mass and residue, which can be expressed as follows:
\begin{eqnarray}
\lambda^2 e^{-\frac{m^2_B}{M^2}}&=&\Pi^{\mathrm{QCD(B)}}_{~\!\not\!{q} }(s_0,M^2),\nonumber\\
	f_{Ps}^2 (\frac{m_{Ps}^2}{m_t+m_{q/Q}})^2 e^{-\frac{m^{2}_{Ps}}{M^{2}}}&=&\Pi^{\mathrm{QCD(Ps)}}(s_0,M^2),\nonumber\\
f_{V}^2 {m_{V}^2} e^{-\frac{m^{2}_V}{M^{2}}} &=&\Pi^{\mathrm{QCD(V)}}_{ g_{\mu\nu}}(s_0,M^2).		
\label{Eq:cor:match1}
\end{eqnarray}
In the subsequent stage of the analysis, the expressions for the masses of the states under consideration are derived by differentiating Eq.~(\ref{Eq:cor:match1}) with respect to $-\frac{1}{M^2}$:
	\begin{eqnarray}
	m^2_{B}=\frac{\frac{d}{d(-\frac{1}{M^2})}\Pi^{\mathrm{QCD(B)}}_{~\not\! q}(s_0,M^2)}{\Pi^{\mathrm{QCD(B)}}_{~\not\!q}(s_0,M^2)},\notag \\ \notag \\
	m^2_{Ps}=\frac{\frac{d}{d(-\frac{1}{M^2})}\Pi^{\mathrm{QCD(Ps)}}(s_0,M^2)}{\Pi^{\mathrm{QCD(S)}}(s_0,M^2)},\notag \\ 
	m^2_{V}=\frac{\frac{d}{d(-\frac{1}{M^2})}\Pi^{\mathrm{QCD(V)}}_{g_{\mu\nu}}(s_0,M^2)}{\Pi^{\mathrm{QCD(V)}}_{g_{\mu\nu}}(s_0,M^2)}.
		\label{Eq:mass:Groundstates1}
	\end{eqnarray}

In the subsequent section, we carry out a detailed numerical evaluation of the QCD sum rules to determine the masses of the states under study. These include the baryons $\Lambda_t$, $\Xi_t$, $\Sigma_t$, $\Xi'_t$, $\Omega_t$, $\Omega_{tcc}$, and $\Omega_{tbb}$, as well as the pseudoscalar and vector mesons containing a single top quark, namely $T_{t\bar{n}}^{\mathrm{Ps}}$, $T_{t\bar{n}}^{\mathrm{V}}$, $T_{t\bar{s}}^{\mathrm{Ps}}$, $T_{t\bar{s}}^{\mathrm{V}}$, $T_{t\bar{c}}^{\mathrm{Ps}}$, $T_{t\bar{c}}^{\mathrm{V}}$, $T_{t\bar{b}}^{\mathrm{Ps}}$, and $T_{t\bar{b}}^{\mathrm{V}}$. We work in the isospin symmetry limit $m_u = m_d$ and denote both up and down quarks by $n$.

\section {QCD SUM-RULE MASS ANALYSIS}\label{sec:three}
In this section, we perform a comprehensive numerical study of the states under consideration, encompassing pseudoscalar and vector mesons, as well as baryons with a single-top quark. The set of input parameters utilized in the analysis is listed in Table~\ref{tab:Parameter}. As stated, nonperturbative condensate contributions are included up to dimension eight. To account for the effects of higher-dimensional terms in the OPE,  we employ the factorization assumption, whereby operators of higher dimension are approximated by combinations of lower-dimensional condensates. This approach effectively incorporates contributions from operators with dimensions beyond six into the calculations The determination of hadron masses in the QCD sum-rule approach relies not only on the underlying physical input parameters but also on several auxiliary quantities, including the mixing parameter $\beta$, the Borel parameter $M^2$, and the continuum threshold $s_0$.

\begin{table}[!htbp]
	\begin{tabular}{|c|c|}
		\hline
		Parameters & Values \\ \hline\hline
			$m_{t}$                                     & $172.56\pm 0.31~\mathrm{GeV}$ \cite{ParticleDataGroup:2022pth}\\ 
		$m_{b}$                                     & $4.78\pm 0.06~\mathrm{GeV}$ \cite{ParticleDataGroup:2022pth}\\
			$m_{c}$                                     & $1.67\pm 0.07~\mathrm{GeV}$ \cite{ParticleDataGroup:2022pth}\\
		$m_{s}$                                   & $93.5 \pm 0.8~\mathrm{MeV}$ \cite{ParticleDataGroup:2022pth}\\
						$m_{d}$                                   & $4.70\pm 0.07~\mathrm{MeV}$ \cite{ParticleDataGroup:2022pth}\\
							$m_{u}$                                   & $2.16\pm 0.07~\mathrm{MeV}$ \cite{ParticleDataGroup:2022pth}\\
		$\langle \bar{q}q \rangle $    & $(-0.24\pm 0.01)^3$ $\mathrm{GeV}^3$ \cite{Belyaev:1982sa}  \\
		$\langle \bar{s}s \rangle $               & $0.8\langle \bar{q}q \rangle$ \cite{Belyaev:1982sa} \\
		$m_{0}^2 $                                & $(0.8\pm0.1)$ $\mathrm{GeV}^2$ \cite{Belyaev:1982sa}\\
		$\langle \overline{q}g_s\sigma Gq\rangle$ & $m_{0}^2\langle \bar{q}q \rangle$ \cite{Belyaev:1982sa}\\
		$\langle g_s^2 G^2 \rangle $              & $4\pi^2 (0.012\pm0.004)$ $~\mathrm{GeV}
		^4 $\cite{Belyaev:1982cd}\\
		$\langle g_s^3 G^3 \rangle $                & $ (0.57\pm0.29)$ $~\mathrm{GeV}^6 $\cite{Narison:2015nxh}\\
		
		\hline\hline
	\end{tabular}%
	\caption{Parameters used in our evaluations.
	}
	\label{tab:Parameter}
\end{table}

   A proper choice of these parameters is essential to guarantee the stability and robustness of the numerical predictions. Within the QCD sum-rule framework, a commonly adopted requirement is that the physical observables demonstrate only a weak dependence on variations of these auxiliary parameters.

\begin{figure}[t]
\centering
\includegraphics[width=0.45\textwidth]{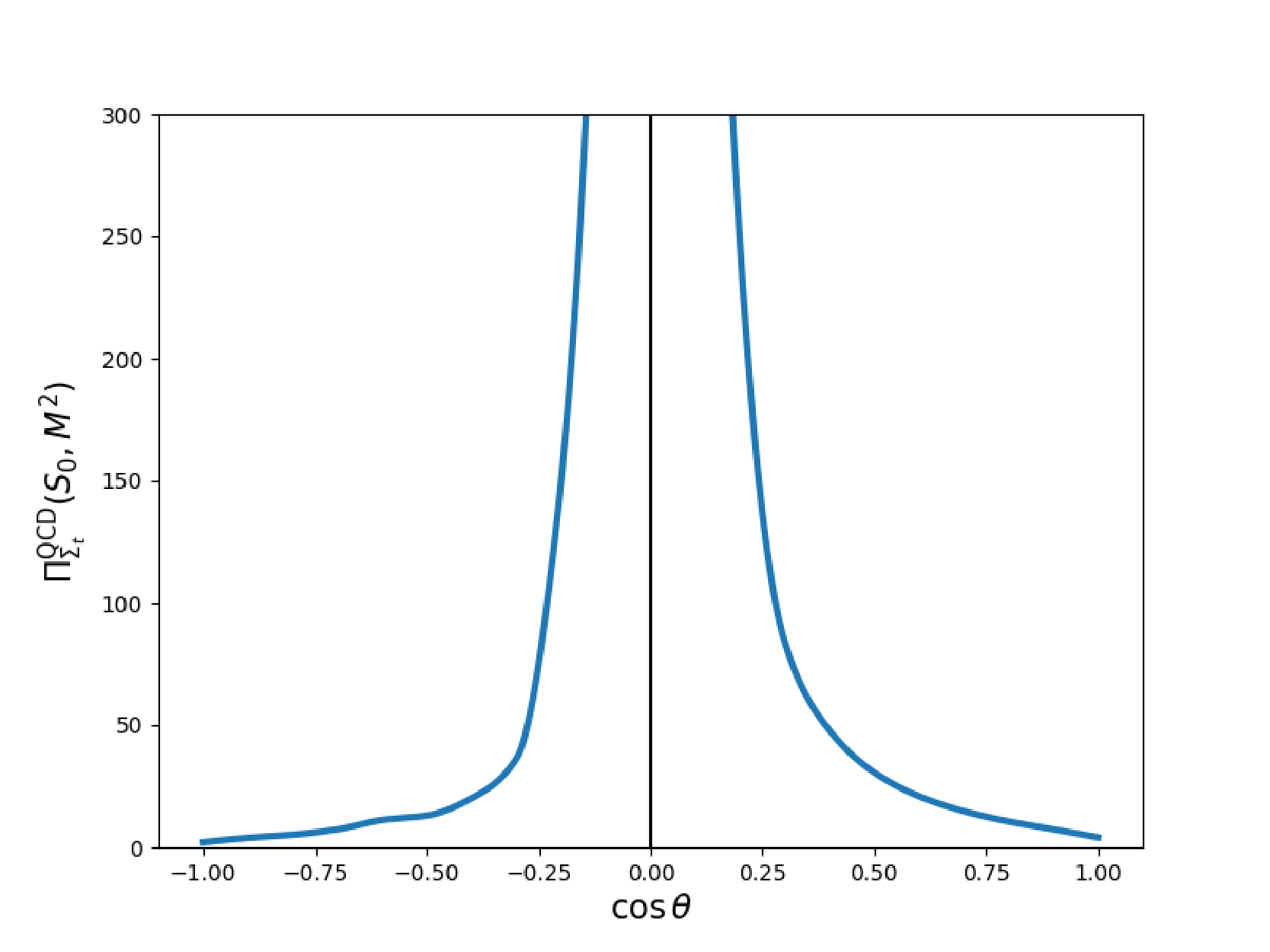}
\caption{$\Pi^{\mathrm{QCD}}_{\Sigma_t}(s_0,M^2)$ (in units of $\mathrm{GeV}^6$) as a function of $\cos\theta$ at the central values of $M^2$ and $s_0$.}
\label{fig:beta}
\end{figure}

The permissible range of the parameter $\beta$ is determined through a detailed analysis based on a parametric representation of the results in terms of $\cos\theta$, where $\beta = \tan\theta$. Stability regions are identified as those intervals in which the results exhibit minimal sensitivity to variations in $\cos\theta$. In our analysis of all single-top baryon  configurations, these regions are found to lie within the interval $-0.61 \leq \beta \leq -0.51$. As a representative example, Fig.~\ref{fig:beta} illustrates the dependence of the QCD correlation function $\Pi^{\mathrm{QCD}}_{\Sigma_t}(s_0,M^2)$ on the mixing parameter $\cos\theta$, evaluated at the central values of the Borel parameter $M^2$ and continuum threshold $s_0$. Similar behavior is observed for the other channels considered in this work.

An additional consistency requirement of the method is that the perturbative contribution be dominant, while nonperturbative corrections are increasingly suppressed with growing operator dimension. In addition, the requirement of ground-state dominance introduces an important restriction, namely that the pole contribution (PC) must be greater than the combined effects of higher resonances and the continuum. Consequently, the upper limit of $M^2$ is determined by imposing the condition that the pole contribution remains larger than the sum of the continuum and excited-state contributions.

Furthermore, the lower bound of the Borel parameter $M^2$ is fixed by imposing the requirement of a reliably convergent operator product expansion. This guarantees that the perturbative contribution remains dominant, while the nonperturbative condensate terms become increasingly suppressed as the operator dimension grows. In practical calculations, these constraints are realized through the following relations:
 
 \begin{eqnarray}
\mathrm{PC}=\frac{\Pi(s_0,M^2)}{\Pi(\infty,M^2)}\geq 0.5,
\end{eqnarray}
and
\begin{equation}
	 \frac{\Pi ^{\mathrm{Dim8}}(s_0,M^2)}{\Pi (s_0,M^2)}\le\ 0.05.
	  	\label{eq:Convergence}
	  \end{equation} 

The continuum threshold, $s_0$, which separates the ground-state contribution from those of excited states and the continuum, is not chosen arbitrarily. After determining the allowed Borel windows using the pole dominance and OPE convergence criteria, the working intervals of $s_0$ are fixed by requiring the extracted hadron masses to exhibit stability against variations of both $s_0$ and the Borel parameter. For the heavy configurations investigated in this work, the continuum threshold is primarily determined by the heavy-quark content and can be parametrized as $\sqrt{s_0}\simeq M_{\rm scale}+\Delta$, where $\Delta$ denotes the energy gap between the ground state and the first excited state. The parameter $\Delta$ is determined independently for each channel by demanding stable QCD sum-rule predictions while simultaneously preserving pole dominance and OPE convergence within the corresponding Borel window. For example, in the $\Lambda_t$ channel, we obtain $\Delta\simeq 3.3\text{--}3.6~\mathrm{GeV}$. A similar analysis is carried out for all channels listed in Tables~II and III, and the resulting ranges of $s_0$ provide stable and reliable predictions for the extracted hadron masses.

\begin{figure}[t]
\centering
\includegraphics[width=0.45\textwidth]{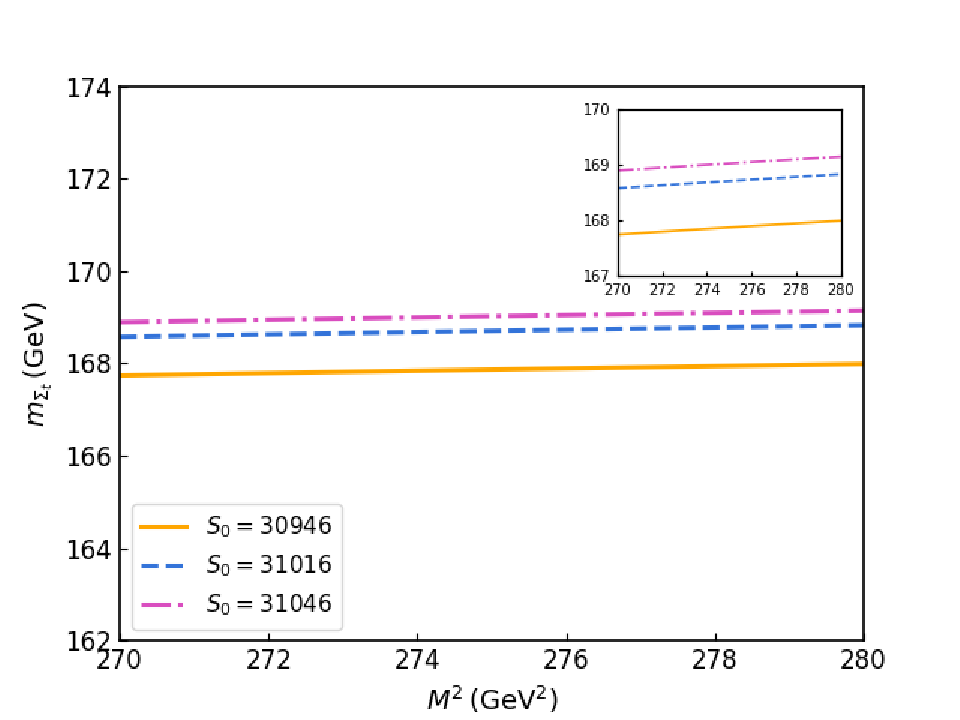}
\includegraphics[width=0.42\textwidth]{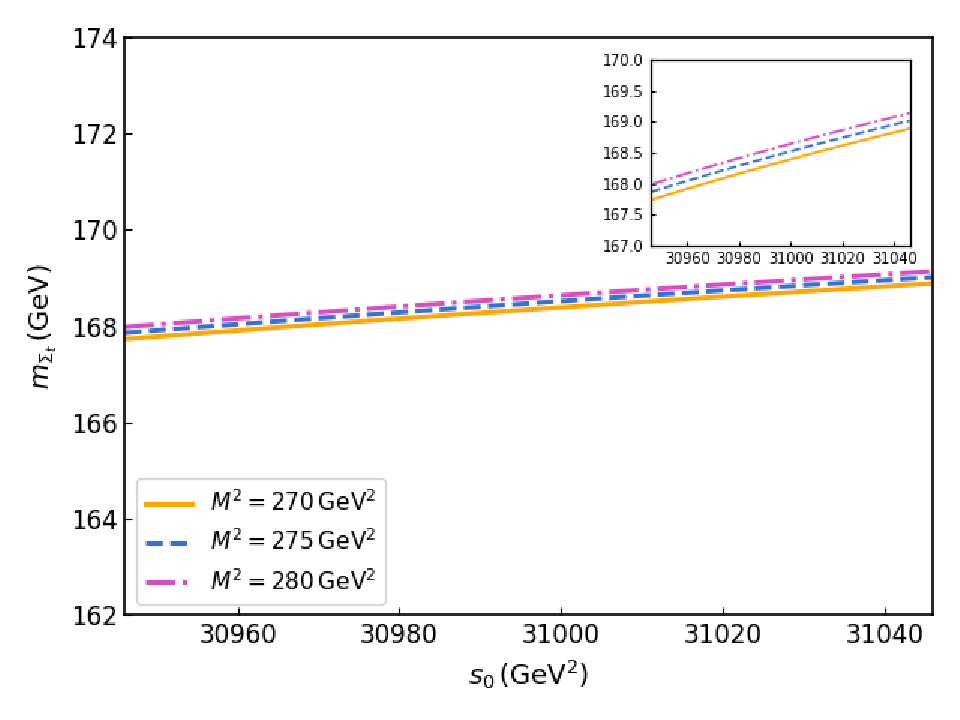}
\caption{Left: Dependence of the extracted $\Sigma_{t}$ mass on the Borel parameter $M^2$ at several fixed values of the continuum threshold $s_0$. 
Right: Dependence of the $\Sigma_{t}$ mass on the continuum threshold $s_0$ for different fixed values of the Borel parameter $M^2$. 
A wider range is displayed for completeness, while the stability region used for the mass extraction is emphasized for clarity.}
\label{fig:mass}
\end{figure}

Our numerical investigation shows that the extracted quantities depend only weakly on variations of these auxiliary parameters within their allowed working regions. As a representative illustration of this behavior, the dependence of the mass on $M^2$ and $s_0$ is shown in Fig.~\ref{fig:mass} for the $\Sigma_t$ baryon. The results clearly exhibit a stable plateau with respect to both the Borel parameter and the continuum threshold over the selected intervals.

After establishing the working regions of the auxiliary parameters, we present the resulting mass predictions for the baryons $\Lambda_t$, $\Xi_t$, $\Sigma_t$, $\Xi'_t$, $\Omega_t^{0}$, $\Omega_{tcc}$, and $\Omega_{tbb}$, as well as pseudoscalar and vector mesons containing a single top quark, namely $T_{t\bar{n}}^{\mathrm{Ps}}$, $T_{t\bar{n}}^{\mathrm{V}}$, $T_{t\bar{s}}^{\mathrm{Ps}}$, $T_{t\bar{s}}^{\mathrm{V}}$, $T_{t\bar{c}}^{\mathrm{Ps}}$, $T_{t\bar{c}}^{\mathrm{V}}$, $T_{t\bar{b}}^{\mathrm{Ps}}$, and $T_{t\bar{b}}^{\mathrm{V}}$, in Tables~\ref{tab:single_charmed_baryons_J12} and \ref{tab:top_mesons}, respectively. The quoted uncertainties are evaluated by systematically incorporating the main sources of error inherent to the sum-rule framework. These include the quark--hadron duality approximation, the remaining dependence on the Borel parameter $M^2$ and the continuum threshold $s_0$, as well as the propagation of uncertainties associated with the input parameters, such as quark masses and the relevant nonperturbative condensates.
\begin{table}[h!]
\centering
\renewcommand{\arraystretch}{1.25}
\begin{tabular}{|c| c| c| c| c |c|}
\hline\hline
Baryon & Quark content  & $M^2~(\mathrm{GeV^2})$  & $s_0 ~(\mathrm{GeV^2})$&Mass $(\mathrm{GeV})$&Mass of quark content  $(\mathrm{GeV})$\\
\hline
$\Lambda_t$   & $tnn$  & $270-280$ & $30946-31046$&$169.2{}^{+1.4}_{-1.6}$&$172.57{}^{+0.31}_{-0.31}$ \\ \hline 
$\Xi_t$       & $tns$  & $270-280$& $30956-31056$&$172.5{}^{+2.5}_{-3.1}$&$172.66{}^{+0.31}_{-0.31}$ \\ 
\hline \hline 
$\Sigma_t$  & $tnn$ & $270-280$& $30946-31046$&$168.7{}^{+1.5}_{-1.6}$& $172.57{}^{+0.31}_{-0.31}$\\ \hline 
$\Xi'_t$  & $tns$  & $270-280$ & $30956-31056$&$173.2{}^{+3.4}_{-3.8}$ &$172.66{}^{+0.31}_{-0.31}$\\ \hline 
$\Omega_t$  & $tss$  & $270-280$ & $30966-31066$&$176.3{}^{+4.8}_{-4.2}$&$172.75{}^{+0.31}_{-0.31}$ \\
\hline
$\Omega_{tcc}$   & $tcc$ & $280-290$ & $31701-31801$&$177.0{}^{+5.2}_{-5.2}$&$175.04{}^{+0.32}_{-0.32}$ \\ \hline 
$\Omega_{tbb}$  & $tbb$  & $290-300$ & $33880-33980$&$182.8{}^{+5.9}_{-5.3}$&$180.86{}^{+0.32}_{-0.32}$ \\
\hline\hline
\end{tabular}
\caption{ Masses of single-top baryons evaluated within the QCD sum-rule framework, using standard Borel stability and continuum threshold criteria.}
\label{tab:single_charmed_baryons_J12}
\end{table}

\begin{table}[h!]
\centering
\renewcommand{\arraystretch}{1.3}
\begin{tabular}{|c| c |c |c |c|  }
\hline\hline
Meson & $M^2~(\mathrm{GeV^2})$  & $s_0 ~(\mathrm{GeV^2})$&Mass $(\mathrm{GeV})$&Mass of quark content  $(\mathrm{GeV})$ \\
\hline
$T_{t\bar{n}}^{\mathrm{Ps}}$     &$265-275$&$30946-31016$&$173.2{}^{+1.5}_{-1.2}$&$172.56{}^{+0.31}_{-0.31}$\\\hline 
$T_{t\bar{n}}^{\mathrm{V}}$     &$265-275$&$30946-31016$&$173.2{}^{+1.5}_{-1.2}$&$172.56{}^{+0.31}_{-0.31}$\\\hline 
$T_{t\bar{s}}^{\mathrm{Ps}}$    &$265-275$&$30956-31026$&$173.3{}^{+2.3}_{-2.3}$&$172.65{}^{+0.31}_{-0.31}$\\\hline 
$T_{t\bar{s}}^{\mathrm{V}}$   &$265-275$&$30956-31026$&$173.3{}^{+2.3}_{-2.3}$&$172.65{}^{+0.31}_{-0.31}$\\\hline 
$T_{t\bar{c}}^{\mathrm{Ps}}$    &$275-285$&$30976-31046$&$175.1{}^{+3.5}_{-3.6}$&$173.77{}^{+0.31}_{-0.31}$\\\hline 
$T_{t\bar{c}}^{\mathrm{V}}$   &$275-285$&$30976-31046$&$175.1{}^{+3.5}_{-3.6}$&$173.77{}^{+0.31}_{-0.31}$\\\hline 
$T_{t\bar{b}}^{\mathrm{Ps}}$   &$285-295$&$31950-32020$&$176.5{}^{+3.7}_{-3.6}$&$176.68{}^{+0.31}_{-0.31}$ \\\hline 
$T_{t\bar{b}}^{\mathrm{V}}$   &$285-295$&$31950-32020$&$176.5{}^{+3.7}_{-3.6}$&$176.68{}^{+0.31}_{-0.31}$ \\
\hline\hline
\end{tabular}
\caption{Hypothetical meson-like configurations formed from a single-top quark bound to either a light or a heavy antiquark.}
\label{tab:top_mesons}
\end{table}

A careful inspection of Tables~\ref{tab:single_charmed_baryons_J12} and \ref{tab:top_mesons} shows that the extracted masses are, within theoretical uncertainties, generally close to the naive sums of the corresponding constituent quark masses. Based on the extracted central masses of our predictions, several of the investigated hypothetical states appear slightly below the corresponding constituent quark thresholds within the intrinsic uncertainties of the QCD sum-rule framework, which may point to possible binding effects. In particular, the inferred binding energies are approximately $3.4~\mathrm{GeV}$, $0.2~\mathrm{GeV}$, and $3.9~\mathrm{GeV}$ for the $\Lambda_t$, $\Xi_t$, and $\Sigma_t$ states, respectively, while similar binding effects of about $0.2~\mathrm{GeV}$ are found for the $T_{t\bar b}^{\mathrm{Ps}}$ and $T_{t\bar b}^{\mathrm{V}}$ states. Moreover, when the full uncertainty ranges, including both upper and lower error bounds, are taken into account in a conservative analysis, several channels remain compatible with weak binding. Positive mass differences are still obtained for $\Lambda_t$ ($5.3~\mathrm{GeV}$), $\Xi_t$ ($3.6~\mathrm{GeV}$), $\Sigma_t$ ($5.8~\mathrm{GeV}$), $\Xi'_t$ ($3.7~\mathrm{GeV}$), $\Omega_t$ ($0.9~\mathrm{GeV}$), $\Omega_{tcc}$ ($3.5~\mathrm{GeV}$), and $\Omega_{tbb}$ ($3.7~\mathrm{GeV}$), as well as for the hypothetical mesons, the smallest effect is found for the $T_{t\bar n}^{\mathrm{Ps}}$ and $T_{t\bar n}^{\mathrm{V}}$ states ($0.9~\mathrm{GeV}$), followed by the $T_{t\bar s}^{\mathrm{Ps}}$ and $T_{t\bar s}^{\mathrm{V}}$ channels ($1.9~\mathrm{GeV}$), and the $T_{t\bar c}^{\mathrm{Ps}}$ and $T_{t\bar c}^{\mathrm{V}}$ states ($2.6~\mathrm{GeV}$). The largest mass separation is observed for the bottom channel, where both $T_{t\bar b}^{\mathrm{Ps}}$ and $T_{t\bar b}^{\mathrm{V}}$ exhibit a mass difference of about $4.1~\mathrm{GeV}$. Although these results do not establish definitive evidence for deeply bound configurations, they indicate that a weakly bound interpretation, particularly in some channels, remains viable once theoretical uncertainties are fully incorporated. These features may be indicative of nontrivial binding dynamics or near-threshold multiquark configurations. 
These results suggest that some of the studied states lie close to, or slightly below, the corresponding effective thresholds, which may indicate the presence of  bound states  or near-threshold multiquark configurations within the QCD sum rule framework. While these patterns naturally reflect the intrinsic structure of the QCD sum rule approach, the overall consistency of the results highlights the robustness and predictive capability of the framework in describing heavy multiquark configurations in a unified manner, thereby providing valuable guidance for future refined theoretical and experimental investigations.

 To further assess the reliability of our results, a direct quantitative comparison between the predicted masses obtained in the present full QCD sum-rule framework and those reported in the HQET sum-rule studies of Refs. \cite{Zhang:2025xxd,Zhang:2025fdp} is presented in Tables  Table~\ref{tab:comparison1} and Table~\ref{tab:comparison2}.
 
 As shown in Table~\ref{tab:comparison1}, our baryon mass predictions are generally consistent with the HQET results. The masses of the $(\Lambda_t)$, $(\Xi_t)$, $(\Sigma_t)$, and $(\Xi’_t)$ baryons are found to be slightly lower than the corresponding HQET values, whereas the predicted mass of the $(\Omega_t)$ baryon is slightly higher. In the meson sector, Table~\ref{tab:comparison2}, an excellent agreement is observed, with the predicted pseudoscalar and vector meson masses differing from the HQET results by only a fraction of a percent.
The remaining small differences can be mainly attributed to three factors. First, unlike HQET, the present analysis is performed in the full QCD framework, where the finite top-quark mass is retained explicitly, naturally incorporating finite-mass effects beyond the infinite-mass approximation. Second, the two approaches employ different interpolating currents and correlation-function structures, leading to different couplings to the physical states. Finally, the treatment of nonperturbative contributions differs, particularly regarding the implementation of higher-dimensional condensates in the operator product expansion. These methodological differences are expected to produce small quantitative shifts while preserving the overall consistency between the two approaches.

\begin{table}[h!]
\centering
\renewcommand{\arraystretch}{1.25}
\begin{tabular}{|c| c| c| }
\hline\hline
Baryon & This \, work& Ref.\cite{Zhang:2025xxd}\\
\hline
$\Lambda_t$   &$169.2{}^{+1.4}_{-1.6}$&$173.63{}^{+0.32}_{-0.33}$ \\ \hline 
$\Xi_t$       &$172.5{}^{+2.5}_{-3.1}$&$173.76{}^{+0.32}_{-0.33}$ \\ 
\hline \hline 
$\Sigma_t$  &$168.7{}^{+1.5}_{-1.6}$& $173.83{}^{+0.32}_{-0.32}$\\ \hline 
$\Xi'_t$  &$173.2{}^{+3.4}_{-3.8}$ &$173.95{}^{+0.32}_{-0.32}$\\ \hline 
$\Omega_t$  &$176.3{}^{+4.8}_{-4.2}$&$174.08{}^{+0.33}_{-0.33}$ \\
\hline\hline
\end{tabular}
\caption{ Comparison of the predicted masses of single-top baryons with previous QCD sum rule predictions.}
\label{tab:comparison1}
\end{table}

\begin{table}[h!]
\centering
\renewcommand{\arraystretch}{1.3}
\begin{tabular}{|c| c |c |  }
\hline\hline
Meson &This \, work& Ref.\cite{Zhang:2025fdp} \\
\hline
$T_{t\bar{n}}^{\mathrm{Ps}}$     &$173.2{}^{+1.5}_{-1.2}$&$173.03{}^{+0.30}_{-0.30}$\\\hline 
$T_{t\bar{n}}^{\mathrm{V}}$     &$173.2{}^{+1.5}_{-1.2}$&$173.03{}^{+0.30}_{-0.30}$\\\hline 
$T_{t\bar{s}}^{\mathrm{Ps}}$    &$173.3{}^{+2.3}_{-2.3}$&$173.12{}^{+0.31}_{-0.30}$\\\hline 
$T_{t\bar{s}}^{\mathrm{V}}$   &$173.3{}^{+2.3}_{-2.3}$&$173.12{}^{+0.31}_{-0.30}$\\
\hline\hline
\end{tabular}
\caption{Comparison of the predicted masses of single-top mesons with previous QCD sum rule predictions.}
\label{tab:comparison2}
\end{table}

\section {Conclusion}\label{sec:four}

Owing to its large mass and extremely short lifetime, the top quark occupies a distinctive role within the Standard Model. Although its rapid decay strongly suppresses conventional hadron formation, it simultaneously makes the top sector a powerful probe of short-distance QCD dynamics and possible physics beyond the Standard Model, providing a unique arena for testing fundamental interactions at the highest energy scales.

In this work, we have carried out a systematic theoretical investigation of baryonic and mesonic states containing a single top quark within the QCD sum-rule framework. Our analysis includes the baryons $\Lambda_t$, $\Xi_t$, $\Sigma_t$, $\Xi'_t$, $\Omega_t$, $\Omega_{tcc}$, and $\Omega_{tbb}$, together with the pseudoscalar and vector mesons $T_{t\bar{n}}^{\mathrm{Ps}}$, $T_{t\bar{n}}^{\mathrm{V}}$, $T_{t\bar{s}}^{\mathrm{Ps}}$, $T_{t\bar{s}}^{\mathrm{V}}$, $T_{t\bar{c}}^{\mathrm{Ps}}$, $T_{t\bar{c}}^{\mathrm{V}}$, $T_{t\bar{b}}^{\mathrm{Ps}}$, and $T_{t\bar{b}}^{\mathrm{V}}$. The resulting spectra provide a systematic benchmark for such hypothetical systems and offer insight into heavy-quark dynamics in an extreme kinematic regime.

Beyond the mass predictions, our analysis indicates that several channels may exhibit nontrivial dynamical features associated with binding. At the level of the extracted central masses, a number of states lie slightly below their corresponding constituent quark thresholds, suggesting possible binding effects in selected channels. Furthermore, when the full uncertainty ranges are taken into account in a conservative manner, additional states become compatible with weakly bound configurations, thereby extending the set of channels for which binding effects cannot be excluded. While these indications do not constitute evidence for stable top hadrons in the conventional sense, they are consistent with the presence of nontrivial attractive QCD dynamics in some channels and provide useful insight into the behavior of strong interactions in systems involving the top quark.

Overall, these results support the view that top-containing hadronic configurations can serve as useful theoretical laboratories for exploring nonperturbative QCD. The predicted spectra may act as reference points for future lattice and continuum studies, and may also guide prospective searches for top-related phenomena in future high-energy experiments. In this context, our findings contribute to ongoing efforts to clarify the role of the top quark in strong interactions and its possible implications for new-physics scenarios.


\begin{thebibliography}{99}

\bibitem{CDF:1995wbb}
F.~Abe \textit{et al.} [CDF],
``Observation of top quark production in $\bar{p}p$ collisions,''
\href{https://doi.org/10.1103/PhysRevLett.74.2626}{Phys. Rev. Lett. \textbf{74}, 2626-2631 (1995)},
\href{https://arxiv.org/pdf/hep-ex/9503002}{[arXiv:hep-ex/9503002 [hep-ex]].}

\bibitem{D0:1995jca}
S.~Abachi \textit{et al.} [D0],
``Observation of the top quark,''
\href{https://doi.org/10.1103/PhysRevLett.74.2632}{Phys. Rev. Lett. \textbf{74}, 2632-2637 (1995)},
\href{https://arxiv.org/pdf/hep-ex/9503003}{[arXiv:hep-ex/9503003 [hep-ex]].}

\bibitem{Bigi:1986jk}
I.~I.~Y.~Bigi, Y.~L.~Dokshitzer, V.~A.~Khoze, J.~H.~Kuhn and P.~M.~Zerwas,
``Production and Decay Properties of Ultraheavy Quarks,''
\href{https://doi.org/10.1016/0370-2693(86)91275-X}{Phys. Lett. B \textbf{181}, 157-163 (1986)},

  
\bibitem{Shibata:2008sy}
A.~Shibata,
``Top Quark Physics at the LHC,''
\href{https://arxiv.org/pdf/0808.0031}{[arXiv:0808.0031 [hep-ex]].}
  
\bibitem{Schrempp:1996fb}
B.~Schrempp and M.~Wimmer,
``Top quark and Higgs boson masses: Interplay between infrared and ultraviolet physics,''
\href{https://doi.org/10.1016/0146-6410(96)00059-2}{Prog. Part. Nucl. Phys. \textbf{37}, 1-90 (1996)},
\href{https://arxiv.org/pdf/hep-ph/9606386}{[arXiv:hep-ph/9606386 [hep-ph]].}
   
\bibitem{Atwood:2000tu}
D.~Atwood, S.~Bar-Shalom, G.~Eilam and A.~Soni,
\href{https://doi.org/10.1016/S0370-1573(00)00112-5}{Phys. Rept. \textbf{347}, 1-222 (2001)},
\href{https://arxiv.org/pdf/hep-ph/0006032}{[arXiv:hep-ph/0006032 [hep-ph]].}
\bibitem{Kehoe:2007px}
R.~Kehoe, M.~Narain and A.~Kumar,
``Review of Top Quark Physics Results,''
\href{https://doi.org/10.1142/S0217751X08039293}{Int. J. Mod. Phys. A \textbf{23}, 353-470 (2008)},
\href{https://arxiv.org/pdf/0712.2733}{[arXiv:0712.2733 [hep-ex]].}

\bibitem{Plehn:2011tg}
T.~Plehn and M.~Spannowsky,
\href{https://doi.org/10.1088/0954-3899/39/8/083001}{J. Phys. G \textbf{39}, 083001 (2012)},
\href{https://arxiv.org/pdf/1112.4441}{[arXiv:1112.4441 [hep-ph]].}

\bibitem{Wagner:2005jh}
W.~Wagner,
``Top quark physics in hadron collisions,''
\href{https://arxiv.org/pdf/hep-ph/0507207}{Rept. Prog. Phys. \textbf{68}, 2409-2494 (2005)},
\href{https://doi.org/10.1088/0034-4885/68/10/R03}{[arXiv:hep-ph/0507207 [hep-ph]].}

\bibitem{Quadt:2006dqn}
A.~Quadt,
``Top Quark Physics at Hadron Colliders,''
\href{https://doi.org/10.1140/epjc/s2006-02631-6}{Eur. Phys. J. C \textbf{48}, 835-1000 (2006)}

\bibitem{Galtieri:2011yd}
A.~B.~Galtieri, F.~Margaroli and I.~Volobouev,
``Precision measurements of the top quark mass from the Tevatron in the pre-LHC era,''
\href{https://doi.org/10.1088/0034-4885/75/5/056201}{Rept. Prog. Phys. \textbf{75}, 056201 (2012)},
\href{https://arxiv.org/pdf/1109.2163}{[arXiv:1109.2163 [hep-ex]].}


\bibitem{Aguilar-Saavedra:2014kpa}
J.~A.~Aguilar-Saavedra, D.~Amidei, A.~Juste and M.~Perez-Victoria,
``Asymmetries in top quark pair production at hadron colliders,''
\href{https://doi.org/10.1103/RevModPhys.87.421}{Rev. Mod. Phys. \textbf{87}, 421-455 (2015)},
\href{https://arxiv.org/pdf/1406.1798}{[arXiv:1406.1798 [hep-ph]].}

\bibitem{Hagiwara:2016rdv}
K.~Hagiwara, K.~Ma and H.~Yokoya,
``Probing CP violation in $e^{+}e^{-}$ production of the Higgs boson and toponia,''
\href{https://doi.org/10.1007/JHEP06(2016)048}{JHEP \textbf{06}, 048 (2016)}
\href{https://arxiv.org/pdf/1602.00684}{[arXiv:1602.00684 [hep-ph]].}

\bibitem{Chen:2016zbz}
L.~Chen, O.~Dekkers, D.~Heisler, W.~Bernreuther and Z.~G.~Si,
``Top-quark pair production at next-to-next-to-leading order QCD in electron positron collisions,''
\href{https://doi.org/10.1007/JHEP12(2016)098}{JHEP \textbf{12}, 098 (2016)},
\href{https://arxiv.org/pdf/1610.07897}{[arXiv:1610.07897 [hep-ph]].}

\bibitem{Agashe:2016bok}
K.~Agashe, R.~Franceschini, D.~Kim and M.~Schulze,
``Top quark mass determination from the energy peaks of b-jets and B-hadrons at NLO QCD,''
\href{https://doi.org/10.1007/JHEP06(2016)048}{Eur. Phys. J. C \textbf{76}, no.11, 636 (2016)},
\href{https://arxiv.org/pdf/1602.00684}{[arXiv:1603.03445 [hep-ph]].}

\bibitem{Kuhn:1987ty}
J.~H.~Kuhn and P.~M.~Zerwas,
``The Toponium Scenario,''
\href{https://doi.org/10.1016/0370-1573(88)90075-0}{Phys. Rept. \textbf{167}, 321 (1988)}

\bibitem{Barger:1987xg}
V.~D.~Barger, E.~W.~N.~Glover, K.~Hikasa, W.~Y.~Keung, M.~G.~Olsson, C.~J.~Suchyta, III and X.~R.~Tata,
``Superheavy Quarkonium Production and Decays: A New Higgs Signal,''
\href{https://doi.org/10.1103/PhysRevD.35.3366}{Phys. Rev. D \textbf{35}, 3366 (1987)}

\bibitem{Fadin:1990wx}
V.~S.~Fadin, V.~A.~Khoze and T.~Sjostrand,
``On the Threshold Behavior of Heavy Top Production,''
\href{}{Z. Phys. C \textbf{48}, 613-622 (1990)},

\bibitem{Sumino:1997ve}
Y.~Sumino,
``Top quark pair production and decay near threshold in e+ e- collisions,''
\href{}{Acta Phys. Polon. B \textbf{28}, 2461-2478 (1997)},
\href{https://arxiv.org/pdf/hep-ph/9711233}{[arXiv:hep-ph/9711233 [hep-ph]].}

\bibitem{Penin:2005eu}
A.~A.~Penin, V.~A.~Smirnov and M.~Steinhauser,
``Heavy quarkonium spectrum and production/annihilation rates to order beta**3(0) alpha**3(s),''
\href{https://doi.org/10.1016/j.nuclphysb.2005.03.028}{Nucl. Phys. B \textbf{716}, 303-318 (2005)},
\href{https://arxiv.org/pdf/hep-ph/0501042}{[arXiv:hep-ph/0501042 [hep-ph]].}

\bibitem{Hagiwara:2008df}
K.~Hagiwara, Y.~Sumino and H.~Yokoya,
``Bound-state Effects on Top Quark Production at Hadron Colliders,''
\href{https://doi.org/10.1016/j.physletb.2008.07.006}{Phys. Lett. B \textbf{666}, 71-76 (2008)},
\href{https://arxiv.org/pdf/0804.1014}{[arXiv:0804.1014 [hep-ph]].}

\bibitem{Kiyo:2008bv}
Y.~Kiyo, J.~H.~Kuhn, S.~Moch, M.~Steinhauser and P.~Uwer,
``Top-quark pair production near threshold at LHC,''
\href{https://doi.org/10.1140/epjc/s10052-009-0892-7}{Eur. Phys. J. C \textbf{60}, 375-386 (2009)},
\href{https://arxiv.org/pdf/0812.0919}{[arXiv:0812.0919 [hep-ph]].}


\bibitem{Beneke:2015kwa}
M.~Beneke, Y.~Kiyo, P.~Marquard, A.~Penin, J.~Piclum and M.~Steinhauser,
``Next-to-Next-to-Next-to-Leading Order QCD Prediction for the Top Antitop $S$-Wave Pair Production Cross Section Near Threshold in $e^+e^-$ Annihilation,''
\href{https://doi.org/10.1103/PhysRevLett.115.192001}{Phys. Rev. Lett. \textbf{115}, no.19, 192001 (2015)},
\href{https://arxiv.org/pdf/1506.06864}{[arXiv:1506.06864 [hep-ph]].}
\bibitem{Fuks:2021xje}
B.~Fuks, K.~Hagiwara, K.~Ma and Y.~J.~Zheng,
``Signatures of toponium formation in LHC run 2 data,''
\href{https://doi.org/10.1103/PhysRevD.104.034023}{Phys. Rev. D \textbf{104}, no.3, 034023 (2021)},
\href{https://arxiv.org/pdf/2102.11281}{[arXiv:2102.11281 [hep-ph]].}

\bibitem{Akbar:2024brg}
N.~Akbar, I.~Asghar and Z.~Ahmad,
``Mass Spectrum, Radii, and Radiative Decay Widths of Toponium,''
\href{https://arxiv.org/pdf/2411.08548}{[arXiv:2411.08548 [hep-ph]].}



\bibitem{Jiang:2024fyw}
S.~J.~Jiang, B.~Q.~Li, G.~Z.~Xu and K.~Y.~Liu,
``Study on Toponium: Spectrum and Associated Processes,''
\href{https://arxiv.org/pdf/2412.18527}{[arXiv:2412.18527 [hep-ph]].}

\bibitem{Fuks:2024yjj}
B.~Fuks, K.~Hagiwara, K.~Ma and Y.~J.~Zheng,
``Simulating toponium formation signals at the LHC,''
\href{https://doi.org/10.1140/epjc/s10052-025-13853-3}{Eur. Phys. J. C \textbf{85}, no.2, 157 (2025)},
\href{https://arxiv.org/pdf/2411.18962}{[arXiv:2411.18962 [hep-ph]].}

\bibitem{Fadin:1987wz}
V.~S.~Fadin and V.~A.~Khoze,
``Threshold Behavior of Heavy Top Production in e+ e- Collisions,''
JETP Lett. \textbf{46}, 525-529 (1987)
LENINGRAD-87-1333.




\bibitem{Strassler:1990nw}
M.~J.~Strassler and M.~E.~Peskin,
``The Heavy top quark threshold: QCD and the Higgs,''
\href{https://doi.org/10.1103/PhysRevD.43.1500}{Phys. Rev. D \textbf{43}, 1500-1514 (1991)}.



\bibitem{Fadin:1994pj}
V.~S.~Fadin, V.~A.~Khoze and M.~I.~Kotsky,
``Top quark polarization as a probe of t anti-t threshold dynamics,''
\href{https://doi.org/10.1007/BF01557234}{Z. Phys. C \textbf{64}, 45-56 (1994)},
\href{https://arxiv.org/pdf/hep-ph/9403246}{[arXiv:hep-ph/9403246 [hep-ph]]}.

\bibitem{Hoang:2000yr}
A.~H.~Hoang, M.~Beneke, K.~Melnikov, T.~Nagano, A.~Ota, A.~A.~Penin, A.~A.~Pivovarov, A.~Signer, V.~A.~Smirnov and Y.~Sumino, \textit{et al.}
``Top - anti-top pair production close to threshold: Synopsis of recent NNLO results,''
\href{https://doi.org/10.1007/s1010500c0003}{Eur. Phys. J. direct \textbf{2}, no.1, 3 (2000)},
\href{https://arxiv.org/pdf/hep-ph/0001286}{[arXiv:hep-ph/0001286 [hep-ph]]}.






\bibitem{Sumino:2010bv}
Y.~Sumino and H.~Yokoya,
``Bound-state effects on kinematical distributions of top quarks at hadron colliders,''
\href{https://doi.org/10.1007/JHEP06(2016)037}{JHEP \textbf{09}, 034 (2010)},
[erratum: JHEP \textbf{06}, 037 (2016)],
\href{https://arxiv.org/pdf/1007.0075}{[arXiv:1007.0075 [hep-ph]]}.



\bibitem{Kawabata:2016aya}
S.~Kawabata and H.~Yokoya,
``Top-quark mass from the diphoton mass spectrum,''
\href{https://doi.org/10.1140/epjc/s10052-017-4884-8}{Eur. Phys. J. C \textbf{77}, no.5, 323 (2017)},
\href{https://arxiv.org/pdf/1607.00990}{[arXiv:1607.00990 [hep-ph]]}.


\bibitem{Reuter:2018rbq}
J.~Reuter, B.~C.~Nejad, A.~Hoang, W.~Kilian, M.~Stahlhofen, T.~Teubner and C.~Weiss,
``Exclusive Top Threshold Matching at Lepton Colliders,''
\href{https://doi.org/10.22323/1.340.0654}{PoS \textbf{ICHEP2018}, 654 (2019)},
\href{https://arxiv.org/pdf/1811.03950}{[arXiv:1811.03950 [hep-ph]]}.


\bibitem{Aguilar-Saavedra:2024mnm}
J.~A.~Aguilar-Saavedra,
``Toponium hunter{\textquoteright}s guide,''
\href{https://doi.org/10.1103/PhysRevD.110.054032}{Phys. Rev. D \textbf{110}, no.5, 054032 (2024)},
\href{https://arxiv.org/pdf/2407.20330}{[arXiv:2407.20330 [hep-ph]]}.


\bibitem{Garzelli:2024uhe}
M.~V.~Garzelli, G.~Limatola, S.~O.~Moch, M.~Steinhauser and O.~Zenaiev,
``Updated predictions for toponium production at the LHC,''
\href{https://doi.org/10.1016/j.physletb.2025.139532}{Phys. Lett. B \textbf{866}, 139532 (2025)},
\href{https://arxiv.org/pdf/2412.16685}{[arXiv:2412.16685 [hep-ph]]}.

\bibitem{Wang:2024hzd}
G.~L.~Wang, T.~F.~Feng and Y.~Q.~Wang,
``Mass spectra and wave functions of toponia,''
\href{https://doi.org/10.1103/PhysRevD.111.096016}{Phys. Rev. D \textbf{111}, no.9, 096016 (2025)},
\href{https://arxiv.org/pdf/2411.17955}{[arXiv:2411.17955 [hep-ph]]}.



\bibitem{Jafari:2025rmm}
A.~Jafari,
``TOP2024: an overview of experimental results,''
\href{https://doi.org/10.21468/SciPostPhysProc.18.029}{SciPost Phys. Proc. \textbf{18}, 029 (2026)},
\href{https://arxiv.org/pdf/2501.16231}{[arXiv:2501.16231 [hep-ex]].}

\bibitem{Francener:2025tor}
R.~Francener, V.~P.~Goncalves and D.~E.~Martins,
``Investigating the exclusive toponium production at the LHC and FCC,''
\href{https://arxiv.org/pdf/2502.03295}{[arXiv:2502.03295 [hep-ph]]}.

\bibitem{dEnterria:2025ecx}
D.~d'Enterria and K.~Kang,
``Exclusive photon-fusion production of even-spin resonances and exotic QED atoms in high-energy hadron collisions,''
\href{https://doi.org/10.1103/rnxl-v6gd}{Phys. Rev. D \textbf{112}, no.11, 116022 (2025)},
\href{https://arxiv.org/pdf/2503.10952}{[arXiv:2503.10952 [hep-ph]].}

\bibitem{CMS:2025kzt}
A.~Hayrapetyan \textit{et al.} [CMS],
``Observation of a pseudoscalar excess at the top quark pair production threshold,''
\href{https://doi.org/10.1088/1361-6633/adf7d3}{Rept. Prog. Phys. \textbf{88}, no.8, 087801 (2025)},
\href{https://arxiv.org/pdf/2503.22382}{[arXiv:2503.22382 [hep-ex]]}.


\bibitem{ATLAS:2025mvr}
 [ATLAS],
``Observation of a cross-section enhancement near the $t\bar{t}$ production threshold in $\sqrt{s}$ =13 TeV pp collisions with the ATLAS detector,''
ATLAS-CONF-2025-008.




\bibitem{Fuks:2025toq}
B.~Fuks, A.~Hossain and J.~Keaveney,
``Statistical Indications of Toponium Formation in Top Quark Pair Production,''
\href{https://arxiv.org/pdf/2511.02040}{[arXiv:2511.02040 [hep-ph]]}.

\bibitem{Biello:2025idh}
C.~Biello, J.~Mazzitelli, C.~Signorile-Signorile, M.~Wiesemann and G.~Zanderighi,
``Progress on NNLO+PS predictions for top-quark pair production and decay,''
\href{https://doi.org/10.22323/1.512.0086}{PoS \textbf{DIS2025}, 086 (2025)}.






\bibitem{Matsuoka:2025jgm}
Y.~Matsuoka,
``Possible mixing between elementary and bound state fields in the $t\bar{t}$ production excess at the LHC,''
\href{https://arxiv.org/pdf/2510.16828}{[arXiv:2510.16828 [hep-ph]]}.

\bibitem{Sjostrand:2025qez}
T.~Sj{\"o}strand,
``On the threshold behaviour of heavy top production,''
\href{https://arxiv.org/pdf/2510.04590}{[arXiv:2510.04590 [hep-ph]]}.

\bibitem{Fuks:2025wtq}
B.~Fuks, K.~Hagiwara, K.~Ma, L.~Munoz-Aillaud and Y.~J.~Zheng,
``Prospects for toponium formation at the LHC in the single-lepton mode,''
\href{https://arxiv.org/pdf/2509.03596}{[arXiv:2509.03596 [hep-ph]]}.

\bibitem{Goncalves:2025hyx}
V.~P.~Goncalves, L.~Santana and B.~D.~Moreira,
``Exclusive vector-toponium photoproduction in hadronic collisions,''
\href{https://arxiv.org/pdf/2508.19879}{[arXiv:2508.19879 [hep-ph]]}.

\bibitem{Zhu:2025ezg}
Z.~Q.~Zhu, C.~Xiong and Y.~J.~Zhang,
``Triple top baryon $\Omega_{ttt}$,''
\href{https://arxiv.org/pdf/2508.19137}{[arXiv:2508.19137 [hep-ph]]}.

\bibitem{Zhang:2025fdp}
S.~W.~Zhang, X.~Luo, H.~M.~Yang and H.~X.~Chen,
``QCD sum rule study of topped mesons within heavy quark effective theory,''
\href{https://doi.org/10.3390/universe11100334}{Universe \textbf{11}, 10 (2025)},
\href{https://arxiv.org/pdf/2508.03422}{[arXiv:2508.03422 [hep-ph]]}.

\bibitem{Lopez:2025kog}
J.~A.~Lopez and C.~Sandoval,
``Quantum Bootstrap Approach to a Non-Relativistic Potential for Quarkonium systems,''
\href{https://arxiv.org/pdf/2508.02916}{[arXiv:2508.02916 [hep-ph]]}.

\bibitem{Thompson:2025cgp}
E.~J.~Thompson,
``Top Quark Bound States in Finite and Holomorphic Quantum Field Theories,''
\href{https://arxiv.org/pdf/2507.16831}{[arXiv:2507.16831 [physics.gen-ph]]}.


\bibitem{Zhang:2025xxd}
S.~W.~Zhang, W.~H.~Tan, X.~Luo and H.~X.~Chen,
``Topped baryons from QCD sum rules,''
\href{https://arxiv.org/pdf/2507.05895}{[arXiv:2507.05895 [hep-ph]]}.

\bibitem{Shao:2025dzw}
H.~S.~Shao and G.~Wang,
``Analytic NNLO transverse-momentum-dependent soft function for heavy quark pair hadroproduction at threshold,''
\href{https://doi.org/10.1007/JHEP10(2025)164}{JHEP \textbf{10}, 164 (2025)},
\href{https://arxiv.org/pdf/2506.23791}{[arXiv:2506.23791 [hep-ph]]}.

\bibitem{Bai:2025buy}
Y.~Bai, T.~K.~Chen and Y.~Yang,
``Toponia at the HL-LHC, CEPC, and FCC-ee,''
\href{https://arxiv.org/pdf/2506.14552}{[arXiv:2506.14552 [hep-ph]]}.

\bibitem{Ellis:2025nkm}
J.~Ellis,
``Personal memories of 50 years of quarkonia,''
\href{https://doi.org/10.1016/j.nuclphysb.2025.117062}{Nucl. Phys. B \textbf{1018}, 117062 (2025)},
\href{https://arxiv.org/pdf/2506.10643}{[arXiv:2506.10643 [hep-ph]]}.

\bibitem{LeYaouanc:2025mpk}
A.~Le Yaouanc and F.~Richard,
``New resonances at LHC,''
\href{https://arxiv.org/pdf/2506.09490}{[arXiv:2506.09490 [hep-ph]]}.

\bibitem{Xiong:2025iwg}
C.~Xiong and Y.~J.~Zhang,
``Probing Yoctosecond Quantum Dynamics in Toponium Formation at Colliders,''
\href{https://arxiv.org/pdf/2507.05703}{[arXiv:2507.05703 [hep-ph]]}.


\bibitem{Fu:2025yft}
J.~H.~Fu, Y.~J.~Li, H.~M.~Yang, Y.~B.~Li, Y.~J.~Zhang and C.~P.~Shen,
``Toponium: The smallest bound state and simplest hadron in quantum mechanics,''
\href{https://doi.org/10.1103/fqc9-k315}{Phys. Rev. D \textbf{111}, no.11, 114020 (2025)},
\href{https://arxiv.org/pdf/2412.11254}{[arXiv:2412.11254 [hep-ph]]}.


\bibitem{Fu:2025zxb}
J.~H.~Fu, Y.~J.~Zhang, G.~Z.~Xu and K.~Y.~Liu,
``Toponium: Implementation of a toponium model in FeynRules,''
\href{https://arxiv.org/pdf/2504.12634}{[arXiv:2504.12634 [hep-ph]]}.


\bibitem{Fuks:2025sxu}
B.~Fuks,
``Toponium physics at the Large Hadron Collider,''
\href{https://arxiv.org/pdf/2505.03869}{[arXiv:2505.03869 [hep-ph]]}.



\bibitem{Afik:2025ejh}
Y.~Afik, F.~Fabbri, M.~Low, L.~Marzola, J.~A.~Aguilar-Saavedra, M.~M.~Altakach, N.~A.~Asbah, Y.~Bai, H.~Banks and A.~J.~Barr, \textit{et al.}
``Quantum information meets high-energy physics: input to the update of the European strategy for particle physics,''
\href{https://doi.org/10.1140/epjp/s13360-025-06752-9}{Eur. Phys. J. Plus \textbf{140}, no.9, 855 (2025)},
\href{https://arxiv.org/pdf/2504.00086}{[arXiv:2504.00086 [hep-ph]]}.

\bibitem{Luo:2025psq}
S.~Q.~Luo, Q.~Huang and X.~Liu,
``The quest for topped hadrons,''
\href{https://arxiv.org/pdf/2508.17646}{[arXiv:2508.17646 [hep-ph]]}.

\bibitem{Jolly:2026gpe}
F.~A.~Jolly [ATLAS and CMS],
``Measurement of spin correlation and entanglement in ATLAS and CMS,''
\href{https://doi.org/10.1016/j.jspc.2026.100358}{J. Subatomic Part. Cosmol. \textbf{5}, 100358 (2026)},
\href{https://arxiv.org/pdf/2601.04649}{[arXiv:2601.04649 [hep-ex]].}

\bibitem{Garzelli:2026jxh}
M.~V.~Garzelli, G.~Limatola, S.~O.~Moch, M.~Steinhauser and O.~Zenaiev,
``Top quark pairs in the threshold at the LHC,''
\href{https://doi.org/10.1016/j.jspc.2026.100368}{J. Subatomic Part. Cosmol. \textbf{5}, 100368 (2026)},

\bibitem{CMS:2026lsc}
A.~Gevorgyan \textit{et al.} [CMS],
``Search for new particles decaying into top quark-antiquark pairs in proton-proton collisions at $\sqrt{s}$ = 13 TeV,''
\href{https://arxiv.org/pdf/2603.23454}{[arXiv:2603.23454 [hep-ex]].}

\bibitem{Najjar:2025bby}
Z.~R.~Najjar and K.~Azizi,
``Masses of Purely Top-Quark Bound States: Toponium and the Triply-Top Baryon,''
\href{https://arxiv.org/pdf/2511.10053}{[arXiv:2511.10053 [hep-ph]].}




\bibitem{Patel:2025gbw}
R.~V.~Patel, M.~Shah, S.~Patel and B.~Pandya,
``Singly heavy omega baryon spectroscopy in the relativistic framework of an independent quark model,''
\href{https://doi.org/10.1103/x8lw-8vfs}{Phys. Rev. D \textbf{112}, no.5, 056025 (2025)},
\href{https://arxiv.org/pdf/2506.23594}{[arXiv:2506.23594 [hep-ph]].}


\bibitem{Li:2023gbo}
Z.~Y.~Li, G.~L.~Yu, Z.~G.~Wang and J.~Z.~Gu,
\href{https://doi.org/10.1140/epjc/s10052-024-12457-7}{Eur. Phys. J. C \textbf{84}, no.2, 106 (2024)},
\href{https://arxiv.org/pdf/2311.08251}{[arXiv:2311.08251 [hep-ph]].}

\bibitem{Li:2025frt}
Z.~Y.~Li, G.~L.~Yu, Z.~G.~Wang, J.~Z.~Gu and H.~T.~Shen,
\href{https://doi.org/10.1088/1674-1137/adf178}{Chin. Phys. \textbf{49}, no.11, 113107 (2025)},
\href{https://arxiv.org/pdf/2503.01237}{[arXiv:2503.01237 [hep-ph]].}

\bibitem{Gayer:2024akw}
L.~Gayer \textit{et al.} [Hadron Spectrum],
``Highly excited B, B$_{s}$ and B$_{c}$ meson spectroscopy from lattice QCD,''
\href{https://doi.org/10.1007/JHEP01(2025)123}{JHEP \textbf{01}, 123 (2025)}
\href{https://arxiv.org/pdf/2408.02126}{[arXiv:2408.02126 [hep-lat]].}

\bibitem{Kaneko:2023kxx}
T.~Kaneko,
\href{https://doi.org/10.22323/1.430.0238}{PoS \textbf{LATTICE2022}, 238 (2023)},
\href{https://arxiv.org/pdf/2304.01618}{[arXiv:2304.01618 [hep-lat]].}

\bibitem{Luo:2025sns}
X.~Luo, S.~W.~Zhang, H.~X.~Chen, A.~Hosaka, N.~Su and H.~M.~Yang,
``A short review on QCD sum rule studies of P-wave single heavy baryons,''
\href{https://arxiv.org/pdf/2510.13013}{[arXiv:2510.13013 [hep-ph]].}

\bibitem{Ni:2023lvx}
R.~H.~Ni, J.~J.~Wu and X.~H.~Zhong,
``Unified unquenched quark model for heavy-light mesons with chiral dynamics,''
\href{https://doi.org/10.1103/PhysRevD.109.116006}{Phys. Rev. D \textbf{109}, no.11, 116006 (2024)},
\href{https://arxiv.org/pdf/2312.04765}{[arXiv:2312.04765 [hep-ph]].}

\bibitem{Fabiano:1997xh}
N.~Fabiano,
``Top mesons,''
\href{https://doi.org/10.1007/s100520050144}{Eur. Phys. J. C \textbf{2}, 345-350 (1998)},
\href{https://arxiv.org/pdf/hep-ph/9704261}{[arXiv:hep-ph/9704261 [hep-ph]].}


\bibitem{LHCb:2017uwr}
R.~Aaij \textit{et al.} [LHCb],
``Observation of five new narrow $\Omega_c^0$ states decaying to $\Xi_c^+ K^-$,''
\href{https://doi.org/10.1103/PhysRevLett.118.182001}{Phys. Rev. Lett. \textbf{118}, no.18, 182001 (2017)},
\href{https://arxiv.org/pdf/1703.04639}{[arXiv:1703.04639 [hep-ex]].}


\bibitem{LHCb:2020iby}
R.~Aaij \textit{et al.} [LHCb],
``Observation of New $\Xi_c^0$ Baryons Decaying to $\Lambda_c^+ K^-$,''
\href{https://doi.org/10.1103/PhysRevLett.124.222001}{Phys. Rev. Lett. \textbf{124}, no.22, 222001 (2020)},
\href{https://arxiv.org/pdf/2003.13649}{[arXiv:2003.13649 [hep-ex]].}



\bibitem{Belle:2013jfq}
A.~Zupanc \textit{et al.} [Belle],
``Measurement of the Branching Fraction $\mathcal B(\Lambda_c^+ \to p K^- \pi^+)$,''
\href{https://doi.org/10.1103/PhysRevLett.113.042002}{Phys. Rev. Lett. \textbf{113}, no.4, 042002 (2014)},
\href{https://arxiv.org/pdf/1312.7826}{[arXiv:1312.7826 [hep-ex]].}

\bibitem{BaBar:2010zpy}
P.~del Amo Sanchez \textit{et al.} [BaBar],
``Observation of new resonances decaying to $D\pi$ and $D^*\pi$ in inclusive $e^+e^-$ collisions near $\sqrt{s}=$10.58 GeV,''
\href{https://doi.org/10.1103/PhysRevD.82.111101}{Phys. Rev. D \textbf{82}, 111101 (2010)},
\href{https://arxiv.org/pdf/1009.2076}{[arXiv:1009.2076 [hep-ex]].}





		\bibitem{Aliev:2009jt}
		T.~M.~Aliev, K.~Azizi and A.~Ozpineci,
		``Radiative Decays of the Heavy Flavored Baryons in Light Cone QCD Sum Rules,''
		\href{https://doi.org/10.1103/PhysRevD.79.056005}{Phys. Rev. D \textbf{79}, 056005 (2009)},
		\href{https://arxiv.org/pdf/0901.0076}{[arXiv:0901.0076 [hep-ph]].}
		
		
		\bibitem{Aliev:2010uy}
		T.~M.~Aliev, K.~Azizi and M.~Savci,
		``Analysis of the $\Lambda_{b}\rightarrow \Lambda \ell^+\ell^- $ decay in QCD,''
		\href{https://doi.org/10.1103/PhysRevD.81.056006}{Phys. Rev. D \textbf{81}, 056006 (2010)},
		\href{https://arxiv.org/pdf/1001.0227}{[arXiv:1001.0227 [hep-ph]].}
		
		\bibitem{Aliev:2012ru}
		T.~M.~Aliev, K.~Azizi and M.~Savci,
		``Doubly Heavy spin-1/2 Baryon Spectrum in QCD,''
		\href{https://doi.org/10.1016/j.nuclphysa.2012.09.009}{Nucl. Phys. A \textbf{895}, 59-70 (2012)},
		\href{https://arxiv.org/pdf/1205.2873}{[arXiv:1205.2873 [hep-ph]].}
		
		\bibitem{Agaev:2016mjb}
		S.~S.~Agaev, K.~Azizi and H.~Sundu,
		``Mass and decay constant of the newly observed exotic $X(5568)$ state,''
		\href{https://doi.org/10.1103/PhysRevD.93.074024}{Phys. Rev. D \textbf{93}, no.7, 074024 (2016)},
		\href{https://arxiv.org/pdf/1602.08642}{[arXiv:1602.08642 [hep-ph]].}
		
		\bibitem{Azizi:2016dhy}
		K.~Azizi, Y.~Sarac and H.~Sundu,
		``Analysis of $P_c^+(4380)$ and $P_c^+(4450)$ as pentaquark states in the molecular picture with QCD sum rules,''
		\href{https://doi.org/10.1103/PhysRevD.95.094016}{Phys. Rev. D \textbf{95}, no.9, 094016 (2017)},
		\href{https://arxiv.org/pdf/1612.07479}{[arXiv:1612.07479 [hep-ph]].}


\bibitem{Bagan:1991sc}
E.~Bagan, M.~Chabab, H.~G.~Dosch and S.~Narison,
``The Heavy baryons Sigma(c) Sigma(b) from QCD spectral sum rules,''
\href{https://doi.org/10.1016/0370-2693(92)90208-L}{Phys. Lett. B \textbf{278}, 367-370 (1992)}


\bibitem{Sundu:2018nxt}
H.~Sundu, S.~S.~Agaev and K.~Azizi,
``New charged resonance $Z_{c}^{-}(4100)$: the spectroscopic parameters and width,''
\href{}{Eur. Phys. J. C \textbf{79}, no.3, 215 (2019)},
\href{https://doi.org/10.1140/epjc/s10052-019-6737-0}{[arXiv:1812.10094 [hep-ph]].}




		

		\bibitem{ParticleDataGroup:2022pth}
		R.~L.~Workman \textit{et al.} [Particle Data Group],
		``Review of Particle Physics,''
		\href{https://doi.org/10.1093/ptep/ptac097}{PTEP \textbf{2022}, 083C01 (2022)}.
		
		
		
		
		\bibitem{Belyaev:1982sa}
		V.~M.~Belyaev and B.~L.~Ioffe,
		``Determination of Baryon and Baryonic Resonance Masses from QCD Sum Rules. 1. Nonstrange Baryons,''
		\href{http://www.jetp.ras.ru/cgi-bin/e/index/e/56/3/p493?a=list}{Sov. Phys. JETP \textbf{56}, 493-501 (1982)}
		ITEP-59-1982.
		

\bibitem{Belyaev:1982cd}
V.~M.~Belyaev and B.~L.~Ioffe,
``Determination of the baryon mass and baryon resonances from the quantum-chromodynamics sum rule. Strange baryons,''
\href{http://www.jetp.ras.ru/cgi-bin/e/index/e/57/4/p716?a=list}{Sov. Phys. JETP \textbf{57}, 716-721 (1983)
ITEP-132-1982.}
		
		
		\bibitem{Narison:2015nxh} 
		S.~Narison,
		``Decay Constants of Heavy-Light Mesons from QCD,''
	\href{https://doi.org/10.1016/j.nuclphysbps.2016.02.030}{	Nucl.\ Part.\ Phys.\ Proc.\  {\bf 270-272}, 143 (2016)},
		\href{https://arxiv.org/pdf/1511.05903}{[arXiv:1511.05903 [hep-ph]].}











\end{thebibliography}
\end{document}